\documentclass[12pt,a4paper,twoside]{article}
\usepackage{epsfig}
\usepackage{amssymb,latexsym,amsfonts}

\newcommand{\be}{\begin{equation}}
\newcommand{\ee}{\end{equation}}
\newcommand{\bd}{\begin{displaymath}}
\newcommand{\ed}{\end{displaymath}}
\newcommand{\ba}{\begin{array}}
\newcommand{\ea}{\end{array}}
\newcommand{\bt}{\begin{tabular}}
\newcommand{\et}{\end{tabular}}

\newcommand{\bea}{\begin{eqnarray}}
\newcommand{\eea}{\end{eqnarray}}
\newcommand{\bean}{\begin{eqnarray*}}
\newcommand{\eean}{\end{eqnarray*}}
\newcommand{\non}{\nonumber}
\newcommand{\mc}{\multicolumn}

\newcommand{\hlf}{\frac{1}{2}}

\newcommand{\cV}{{\cal V}}
\newcommand{\cF}{{\cal F}}
\newcommand{\cA}{{\cal A}}
\newcommand{\cM}{{\cal M}}
\newcommand{\tD}{\widetilde{D}}
\newcommand{\dif}{\mathrm{d}}

\newcommand{\inp}[2]{\langle #1, #2 \rangle}

\newcommand{\Z}{\mathbb{Z}}

\newcommand{\R}{\mathbb{R}}
\newcommand{\C}{\mathbb{C}}
\newcommand{\h}{\mathbb{H}}

\newcommand{\ov}[1]{\overline{#1}}
\newcommand{\df}[1]{{} * \! #1}

\begin{document}
\begin{titlepage}

\begin{center} 
\today \hfill                  hep-th/0210178

\hfill VUB/TENA/02/06, PAR-LPTHE-02-38, LPTENS-02/40

\vskip 2 cm
{\Large \bf The group theory of oxidation}\\
\vskip 1.25 cm
{Arjan Keurentjes\footnote{email address: Arjan@tena4.vub.ac.be}}\\
\vskip 0.5cm
{\sl Theoretische Natuurkunde, Vrije Universiteit Brussel,\\ Pleinlaan
  2, B-1050 Brussels, Belgium \\}
\end{center}
\vskip 2 cm
\begin{abstract}
\baselineskip=18pt
Dimensional reduction of (super-)gravity theories to 3 dimensions results 
in sigma models on coset spaces $G/H$, such as the $E_8/SO(16)$ coset
in the bosonic sector of 3 dimensional maximal supergravity. The
reverse process, oxidation, is the reconstruction of a higher
dimensional gravity theory from a coset sigma model. Using the group
$G$ as starting point, the higher dimensional models follow
essentially from decomposition into subgroups. All equations of motion
and Bianchi identities can be directly reconstructed from the group
lattice, Kaluza-Klein modifications and Chern-Simons terms are encoded
in the group structure. Manipulations of extended Dynkin diagrams
encode matter content, and (string) dualities. The reflection symmetry
of the ``magic triangle'' for $E_n$ gravities, and approximate reflection
symmetry of the older ``magic triangle'' of supergravities in 4
dimensions, are easily understood in this framework. 
\end{abstract}

\end{titlepage}
\section{Introduction}

It has been appreciated for a long time that supergravity theories, or
more generally, any theory of which general relativity is a subsector,
are rich in algebraic structures. The best established of these are
the coset symmetries that occur in theories that are related to higher
dimensional theories by toroidal compactification, and subsequently
truncating to the massless sector\footnote{In this paper, we will
  refer to this procedure as ``dimensional reduction'', even though
  one can of course reduce over other manifolds than tori.}. The
cosets are of the form $G/H$, where $G$ is a non-compact Lie group,
and $H$ its maximal compact subgroup. For the maximal supergravity
\cite{Cremmer:1978km} theories, the groups $G$ are members of the
exceptional $E_{n(n)}$-series as found by Cremmer and Julia
\cite{Cremmer:1979up}, while for theories with 16 supersymmetries the
groups $SO(8,8+r)$ are found. 

Other papers have found suggestive relations with, and/or made
conjectures involving affine, hyperbolic or still more general Kac-Moody
algebra's (an incomplete list being \cite{Julia:1982gx,
  Nicolai:1988jb, Nicolai:kx, Mizoguchi:1997si, Damour:2001sa,
  West:2001as}), deformed superalgebra's \cite{Cremmer:1998px} and
Borcherds superalgebra's \cite{Henry-Labordere:2002dk}. At the moment,
these structures are only partly understood.   

The supergravities turn up in the low energy description of string
theories, and their hypothetical non-perturbative ancestor,
M-theory. An exciting conjecture suggests that discrete subgroups of
the Cremmer-Julia groups are realized as exact symmetries in
toroidally compactified M-theory \cite{Hull:1994ys, Witten:1995ex}(for
a review see \cite{Obers:1998fb}). Even bolder conjectures have been
put forward pointing out that these discrete subgroups might fit in a
bigger structure, related to the infinite-dimensional
algebra $E_{11}$ and suggesting that some form of $E_{11}$ is realized
as a symmetry in an appropriate, to be discovered, formulation of
M-theory \cite{West:2001as}. 

A lot of the structure is not specific for supersymmetric theories,
and we will not restrict to these. In this paper we study the well
established coset theories. Our starting point will be 3
dimensional theories, where the groups involved are
finite-dimensional. As demonstrated in \cite{Breitenlohner:1987dg,
  Cremmer:1999du}, it is possible to interpret these theories as
dimensional reductions of higher dimensional theories. Reconstructing
the higher dimensional theory from the lower dimensional one is called
``oxidation''. An interesting aspect of oxidation is that, in contrast
to dimensional reduction, it is not unique: there can be different
``branches'' leading to the same lower dimensional theory. In relevant
cases, this nicely coincides with T-duality symmetries from string
theory.  

This paper describes some group theoretical aspects of
the oxidation procedure. Not only does this give a unified point of
view on the case-by-case studies of \cite{Breitenlohner:1987dg,
  Cremmer:1999du}, it also provides insights in the way in which
certain symmetries leave their mark in higher dimensions. Our analysis
puts  some ``observations'' in the literature on firmer ground, and,
under some reasonable assumptions, it is possible to enumerate all the
possible theories that can be oxidized from a certain coset theory. As a
byproduct, the discrete symmetries relating \emph{different}
gravity theories, grouped in so-called ``magic triangles'' can be explained.  

In this paper we will mostly restrict ourselves (as reference
\cite{Cremmer:1999du} also did) to the maximally non-compact, or split
forms of simple Lie groups. In a follow-up paper \cite{Arjannew} (see
also \cite{HJK}) the
formalism will be extended to cover the other
non-compact forms of the various simple Lie-groups. 

The outline of this paper will be as follows. In section \ref{idea} we
will describe general ideas. As an immediate corollary, we find a
first criterion on the subgroups we are looking for. In section
\ref{reduc} we review dimensional reduction. In section \ref{oxi} we
develop a group theoretical recipe for oxidation, compare it to what
is obtained by dimensional reduction, and comment on equations of
motion, Bianchi identities, Kaluza-Klein modifications and
Chern-Simons terms, which all follow from group
theory. Section \ref{diag} is devoted to the graphical language
provided by extended Dynkin diagrams: These encode the matter content
of the theory, and various dualities. Section \ref{triang} explains
the symmetries of two ``magic'' triangles found in the
literature. Section \ref{concl} summarizes, and comments on
 some previous developments in the understanding of symmetries of
 (compactified) theories of gravity. Our conventions on Lie
algebras are summarized in appendix \ref{Lie}. Appendix \ref{coset} lists
relevant subgroup decompositions for the split forms of the simple
finite dimensional Lie groups. With these, and the formalism developed
in the text, the reader can rederive all the results of
\cite{Cremmer:1999du}, and will be able to fill in some small
omissions in that paper (that were mostly filled in in
\cite{Henry-Labordere:2002dk}). Appendix \ref{details} provides more
relevant group theoretic information, included for easy reference.  

We have listed the most relevant papers in the reference section. Many
older papers can be found in \cite{Salam:fm}, while reviews
such as \cite{Obers:1998fb} and \cite{Fre:2001jd} list more recent
references.

We conclude the introduction with a caveat (and an apology) to the
mathematically inclined reader. As common in the physics literature,
we will mention various groups while completely neglecting their
global structure. As an example, we will not distinguish between
$SO(5)$ and $Sp(2)$, but rather use the first form when the group acts
on a ``real'' structure, and the second if it acts on a ``symplectic''
structure. Appendix \ref{details} includes a list of groups,
isomorphic up to a quotient by a subgroup of the center. 

\section{Intuitive idea}\label{idea}

\subsection{Symmetries in reduced gravity theories}

This section reviews various, sometimes old, ideas on the coset
theories resulting from compactifying gravity. Much of the discussion
here can be found in \cite{Julia:1980gr}, whereas the discussion on
the representation of various forms follows \cite{Julia:1997cy} to
some extent. 

In gravity- and coset theories the concept of a vielbein, and its
generalizations play an important role. A vielbein can be regarded as
a local frame. A priori, this associates an element of $GL(D,\R)$ to
each point. Vielbeins have
two indices, a ``curved'' and a ``flat'' one. The curved one
transforms under general coordinate transformations. An element of
$GL(D,\R)$ can be written as the product of a matrix proportional to
the identity, times an element of $SL(D,\R)$ (modulo a $\Z_2$ ambiguity
when $n$ is even). In the following, this diagonal part decouples from
the discussion, and we will further ignore it. 

The flat index transforms under the transformations that takes
orthonormal frames to orthonormal frames, hence $SO(D\!-\!1,1)$. We do not
want to distinguish frames related by orthogonal transformations, so
we divide by these. The vielbein therefore describes the coset
$SL(D,\R)/SO(D\!-\!1,1)$, which can be parametrized by
$$
(D^2-1) - \frac{D(D-1)}{2} = \frac{D(D+1)}{2} - 1
$$
parameters, which is the number of components of a symmetric tensor
minus its trace: the metric. In general relativity, the metric becomes
dynamical, describing the dynamics of a massless field, and the 
degrees of freedom for a massless field should organize in representations
of $SO(D \! - \! 2)$. Therefore, the cosets $SL(D \! - \! 2)/SO(D \! -
\! 2)$ intuitively seem to represent the vielbein formalism restricted
to the degrees of freedom. Moreover, upon dimensional reduction, the
number of degrees of freedom stays constant, and the coset also seems
to be the relevant structure for the lower dimensional theories
\cite{Julia:1980gr}.  

The group $SL(n,\R)$ is easily constructed from $SU(n)$. $SU(n)$ has a
$\Z_2$ automorphism (inner for $n=2$, outer for $n > 2$), that acts as
complex conjugation. Of the antihermitean generators of $SU(n)$, the
imaginary ones change sign, whereas the real ones are
invariant. Multiplying the imaginary generators with $i$, they turn
into real (but hermitian, and therefore symmetric) generators, and the
resulting set of generators generates $SL(n,\R)$. On this real form
the previous outer automorphism now acts on the algebra as transposing
plus multiplication with $-1$, which leaves the anti-symmetric
generators (generating the compact subgroup) invariant. 

There may be other massless fields in the theory, which are classified in
representations of $SL(n,\R)$. Applying an element of $SL(n,\R)/SO(n)$
allows one to switch between $SL(n,\R)$ and $SO(n)$. However, after
conversion to $SL(n,\R)$, fields related by Poincar\'e
duality transform differently. To be precise, Poincare
duality precisely corresponds to the automorphism that inverts the
generators of the non-compact part of the group.

The completely antisymmetric $k$-tensor of $SL(n,\R)$ has ${n \choose
  k}$ components. Poincar\'e duality relates it to the antisymmetric
  $(n-k)$ tensor, which has ${n \choose {n-k}} = {n \choose k}$
  components.  The two have equal dimension, but transform oppositely
  under the non-compact part of $SL(n,\R)$. The various
  representations of $SL(n,\R)$ will be denoted by their
dimension, with a bar over the number whenever the irrep transforms
oppositely under the compact generators. This conforms with notation
used commonly for representations of $SU(n)$. 

Non-trivial representations of $SL(n,\R)$ are not invariant under its
 $\Z_2$ automorphism. There are
 two possibilities: the automorphism maps the representation to an
 equivalent one, or to a different representation. An example of the
 first kind is the adjoint representation, $\bf{n^2-1}$. Its invariant
 part corresponds to the adjoint of $SO(n)$, and its non-invariant
 part is a real symmetric traceless matrix, identified with the degrees
 of freedom of the metric. All other massless bosonic fields are
 forms, completely antisymmetric tensors. The $\Z_2$ automorphism maps
 the $k$-form $\mathbf{n \choose k}$ degrees of freedom to the
 $(n-k)$-tensor $\mathbf{\ov{n \choose k}}$, which is identifies with
 its Poincar\'e dual. 

The form and its dual can only represent the same
 degrees of freedom after projection on the $SO$ subgroup of
 $SL(n,\R)$. It is useful to think of the tensor and its dual
 as a priori independent. Projecting to $SO(n,\R)$ one has to make a
 ``choice of gauge'', and choose either the tensor or its dual to
 represent the relevant degrees of freedom. The other tensor is
 expressed in terms of the first, by means of the antisymmetric
 Levi-Civita tensor. In \cite{Cremmer:1998px} the treatment of tensor
 and dual as independent is taken quite far; their ``twisted
 self-duality constraint'' is analogous to what we have called a
 choice of gauge. Later in this paper, we will associate a Bianchi
 identity to the tensor  field, and the equation of motion to its
 dual. That there is no elementary distinction between the
 elementary and the dual form is our version of ``twisted
 self-duality'' (sometimes called a ``silver rule'' \cite{Julia:1997cy}). 

Reducing a $\tD$-dimensional theory to 3 dimensions, we expect at
least the $SL(\tD\!-\!3)$ symmetry of the $\tD-3$-torus. In fact the
symmetry is always larger. Vectors are dual to scalars in 3
dimensions, and one finds a coset model with the full $SL(\tD\!-\!2)$
symmetry\cite{Cremmer:1998px} from the transverse degrees of
freedom. The symmetry can however still be larger. 

In a 3-dimensional theory based on a coset $G/H$, we expect to find an
  $SL(D \! - \! 2)$ subgroup in\footnote{Although in this paper we
  will restrict to  $G$'s that are split, this statement and the
  following ones are key   ingredients for the general case
  \cite{Arjannew}.} $G$. Decomposing with respect to this subgroup,
  the adjoint representation of $G$ decomposes into the adjoint of
  $SL(D \! - \! 2)$, plus a number of other representations. If $SL(D
  \! - \! 2)$ is interpreted as the symmetry of the transverse degrees
  of freedom, the representations should be interpreted as matter
  fields. There are a few noteworthy points:   
\begin{itemize}
\item The centralizer of $SL(D \! - \! 2)$ in $G$ acts as a symmetry group on
 the $SL(D \! - \! 2)$ representations. We will call the centralizer the
 \emph{U-duality group}, even though most cosets do not have a direct
 relation with string theory (generically they do not even allow a
 supersymmetric extension); 
\item Concentrating on the massless sectors, the
  irreducible representations (irreps) of $SL(D \! - \! 2)$ in the
  decomposition besides the adjoint should allow an
  interpretation as form fields. This leads to a
  constraint that is explained in the next subsection; 
\item In decomposing the adjoint representation, one finds
  self-conjugate, or pairs of conjugate irreps. As conjugation
  corresponds to Poincar\'e duality, a pair of conjugate
  representations is interpreted as a field and its dual;
\item Representations that are selfconjugate under $SL(\tD\!-\!2)$ require
  some care. There are two cases: In each there are only half
  the number of degrees of freedom. For $\tD-2 =4k$ the antisymmetric
  $(\tD/2-1)$-tensors can be (anti-)self-dual, and hence are interpreted
  as (anti-)selfdual form fields; for $\tD = 4k$, half of the fields
  are interpreted as the duals of the other half, but the forms are not
  identified with their duals\footnote{If we had discussed $SU(\tD-2)$
  instead of $SL(\tD-2)$, the corresponding representations would be
  called real (for $\tD-2=4k$) and pseudoreal (for $\tD =4k$)
  respectively.} (see also \cite{Julia:1997cy}). In the latter case
  only forms plus their duals fill complete U-duality
  representations (as an example: The 28 vector fields in maximal
  supergravity in 4 dimensions \cite{Cremmer:1979up} fill half of the
  $\mathbf{56}$ of  $E_{7(7)}$; their duals fill the other half). 
\end{itemize}

Summarizing, the claim is that higher dimensional theories can be recovered
from 3 dimensions by decomposing the 3 dimensional U-duality
group into an $SL(D \! - \! 2)$ group encoding gravity, times a group which
is the relevant U-duality group in the higher dimension. The
irreducible representations of $SL(D \! - \! 2)$ are interpreted as scalars
(singlets), forms (anti-symmetric tensors) and the graviton (the
adjoint irrep). The rest of this paper will be devoted to making this
idea more precise.

\subsection{Regular subgroups of long roots}

There is an immediate consequence of the previous ideas. To make sense
of the recipe, we demand that the only irreps of $SL(D \! - \! 2)$
that are found in the decomposition can be interpreted as one (and
exactly one) graviton, a number of antisymmetric tensors (form
fields), and singlets (scalars). This can be translated to a
constraint on the relevant subgroups. In this section we will make use
of some results in group theory, which can mostly be found in a classical
paper by Dynkin \cite{Dynkin:um}.   

Let ${\cal L}_G$ be the complexification of a simple Lie algebra, and
${\cal L}_{\tilde{G}}$ a subalgebra of ${\cal L}_G$. If ${\cal L}_G$
would be the complexification of $SL(2)$, $A_1$, the discussion below
would trivialize, so we assume that this is not the case. The Cartan
subalgebra of ${\cal L}_{\tilde{G}}$ can be chosen such
that it is embedded in the Cartan subalgebra\footnote{This is true for
  compact forms, and for complexified algebra's, but not
  in general. The discussion on the representations in this
  subsection does however not at all rely on the particular real form
  of the algebra. Hence, to avoid unnecessary complications, we have
  turned to the complexification. I thank B.~Julia for
  pointing out that an earlier version of this subsection was
  inaccurate at this point.} of the algebra ${\cal
  L}_G$. An immediate corollary is that the roots of ${\cal
  L}_{\tilde{G}}$ can be expressed as linear combinations of simple
roots of ${\cal L}_G$. Moreover, the coefficients appearing in the
expansion of the roots of ${\cal  L}_{\tilde{G}}$ in simple roots of
${\cal L}_G$ are integers. 

The Killing form provides a symmetric bilinear two-form on the root
space of ${\cal L}_G$. It is unique up to normalization; we normalize it
such that the long roots of the algebra have length $\sqrt{2}$,
and denote the resulting form as $\inp{ \ \cdot \ }{ \  \cdot \
}_G$. Completely analogously there exists a bilinear form on the root
space of ${\cal L}_{\tilde{G}}$. We normalize this such that
\emph{the long roots of ${\cal L}_{\tilde{G}}$} have length
$\sqrt{2}$, and denote the resulting form as $\langle\ \cdot\ ,\ \cdot \
\rangle_{\tilde{G}}$. Then on ${\cal L}_{\tilde{G}}$ we have defined
two forms, and uniqueness up to normalization implies
\be
\kappa \inp{\ \cdot\ }{\ \cdot \ }_G= \inp{\ \cdot\ }{\ \cdot \
}_{\tilde{G}} 
\ee
where $\kappa$ is an constant, that is actually a positive integer
\cite{Dynkin:um}, and called the index of the subgroup. Obviously, we
can then extend $\langle\ \cdot\ ,\ \cdot \ \rangle_{\tilde{G}}$ to
${\cal L}_G$. 

Consider a root $\alpha$ of ${\cal L}_{\tilde{G}}$. Via standard
 arguments,  $\alpha$, $e_{\alpha}$ and $e_{-\alpha}$ generate an $A_1$
 subalgebra of ${\cal L}_{\tilde{G}}$. Decomposing ${\cal L}_{G}$ with
 respect to this subalgebra, we need to find all weights. If $\beta$
 is a root of ${\cal L}_G$, then the corresponding weight of the $A_1$
 subalgebra is  
\bd
\inp{\alpha}{\beta}_{\tilde{G}}.
\ed
Now take ${\cal L}_{\tilde{G}}$ to be a
 $A_{D-3}$ algebra (the complexification of the algebra of $SL(D \! -
 \! 2)$). It is
 clear that if $\beta \neq \pm \alpha$, but instead  corresponds to a
 weight of the adjoint or an antisymmetric tensor 
 representation, that $\beta$ must correspond to either a zero-weight,
 or a weight of the 2 dimensional irrep of $A_1$. Hence we should
 require that 
\be \label{indexreq}
|\inp{\alpha}{\beta}_{\tilde{G}}| = \kappa |\inp{\alpha}{\beta}_{G}| < 2
\ee
 We have chosen $\beta$ to lie on the root lattice of ${\cal L}_G$, and
 according to \cite{Dynkin:um}, $\alpha$ lies also on the root
 lattice. As we assumed ${\cal L}_G \neq A_1$ and ${\cal L}_G$ simple,
 one can always  find a 
 $\beta$ for which $\inp{\alpha}{\beta}_G \neq 0$. If we also require
 $\beta \neq \pm \alpha$, and the inequality (\ref{indexreq}),
 $\kappa$ must equal one.

Index 1 subgroups are special. They must be
\emph{regular}, which means that that the root lattice of the subgroup
can be chosen to be a sublattice of the lattice of the original
group. All regular subgroups of a given group can be found by a
procedure described by Dynkin \cite{Dynkin:um}, which we will explain in
section \ref{diag}.  

If the group is simply laced, all regular embeddings have index 1. If
the group is non-simply laced, then regular subgroups involving only
short roots are possible. Such subgroups have an index bigger than
one, and hence are excluded.

The conclusion is that regular $SL$ subgroups of long roots are
the appropriate mathematical structure for encoding the graviton in
the theories under study. Regularity of subgroups is a criterion that
has been observed before \cite{Julia:1980gr}, and is implicit in many
discussions on algebraic aspects of compactified gravity. Here we
have justified regularity as a consequence of reasonable
assumptions, instead of observing it to be obeyed by the available data.

The requirement of long roots is correlated
to the observation in the literature that what is referred to
as group disintegration has to start at one end of the Dynkin
diagram \cite{Julia:1982gx}. We postpone a more detailed discussion to
section \ref{diag}.

\section{Decreasing the number of dimensions: Reduction}\label{reduc}

The claim, to be made precise in the next 2 sections, is that the
oxidation process is completely determined by group theory. We will
now review a systematic procedure for dimensional reduction developed
in \cite{Lu:1995yn, Cremmer:1997ct, Cremmer:1999du}, and emphasize
elements that are significant to our discussion. In the next section
we will develop our recipe for oxidation, and demonstrate that it is
precisely inverse to dimensional reduction. 

\subsection{Dimensional reduction}

In \cite{Lu:1995yn, Cremmer:1997ct, Cremmer:1999du} dimensional
reduction is developed as an inductive scheme, by first reducing over
1 dimension, then over a second etc. We will use the possibility of
rotations on the scalar manifold to simplify some expressions. We will
not use some symbols defined in \cite{Lu:1995yn, Cremmer:1997ct,
  Cremmer:1999du}, and on the other hand introduce symbols that do
have a direct significance to our subsequent discussion. 

We reduce a theory in $\tD$ dimensions, over $n$ toroidal
dimensions, to a $D$ dimensional theory. Of course $\tD=n+D$, but we will
keep all symbols to simplify expressions. First we introduce some
definitions. Consider the vector space $\R^n$, with a basis of unit
vectors $e_i$. For explicit computations one can choose $(e_i)^k
= \delta_i^k$. First define 
\be
S=\sum_{i=1}^n e_i
\ee
In the dimensional reduction procedure, we will extensively need the
vectors 
\be
s= \frac{S}{\sqrt{(\tD-2)(D - 2)}} \qquad f_i =
\sqrt{\frac{\tD-2}{D-2}} \left(\frac{S}{n}\right) + f_i^\perp 
\ee
where in turn 
\be
f_i^{\perp} = e_i - \frac{S}{n}  \qquad \rightarrow \qquad \inp{s}{
  f_i^{\perp}}= 0 
\ee  

These vectors obey
\be \label{relations}
\inp{s}{s} = \frac{n}{(\tD-2)(D-2)} \qquad \inp{s}{f_i} =
\frac{1}{D-2} \qquad \inp{f_i}{f_j} = \delta_{ij} + \frac{1}{D-2}. 
\ee
Up to a factor of $2$, these relations are also satisfied by the
vectors $s$ and $f_i$ defined in\cite{Cremmer:1999du}, so the
vectors appearing here may be identified with the vectors in
\cite{Cremmer:1999du} (up to rescaling with $\sqrt{2}$ which we absorb
in the dilaton), as they can be rotated into each other by an
$O(n)$ rotation.  

The reason for these conventions is that the groups $SL(n)$ (their
discrete versions $SL(n,\Z)$ describing the symmetry group of the
compactification torus) play a prominent role. The $f_j^{\perp}$ form
a non-orthogonal basis of the $(n-1)$ dimensional subspace of $\R^n$
orthogonal to $s$. There are $n$ vectors $f_j^{\perp}$, so this basis
is overcomplete. The combinations
\be
f_i - f_j = f_i^{\perp} - f_j^{\perp} = e_i - e_j
\ee
span an $(n-1)$-dimensional lattice; this is a well-known
representation of the root lattice of $SL(n)$ (\cite{Cremmer:1999du}
defines the Dynkin diagram with the above vectors). The
$f_{j}^{\perp}$ are now easily interpreted: From the innerproducts of the
root vectors of $SL(n)$ with the $f_{j}^{\perp}$ one finds that they
span the weight lattice of $SL(n)$. The $n$ $f_j^{\perp}$ form the
weights of the $n$-dimensional, fundamental representation. The
weights encode the possible eigenvalues of the generators, and the relation 
$\sum_j f_j^{\perp}=0$ expresses tracelessness of the generators of the
algebra of the $n$-dimensional representation.

The Kaluza-Klein metric ansatz is
\be \label{met}
\dif s_{\tD}^2 = e^{\inp{s}{\phi}} \dif s_{D}^2 + e^{-\frac{D-2}{n}
  \inp{s}{\phi}}\sum_{i=1}^{n} e^{-\inp{f_i^{\perp}}{\phi}} (h_i)^2 
\ee
with
\be
h_j = \tilde{\gamma}^i_{\phantom{i}j}(\dif z_i + \hat\cA_i),\qquad
\tilde{\gamma}^i_{\phantom{i}j} = \delta^i_{\phantom{i}j} +
\cA^i_{(0)j} 
\ee
The matrix $\cA^i_{(0)j}$ has non-zero entries only for $i > j$. In
spite of its different appearance the metric eq.(\ref{met}) is
nothing but a rewriting of eq.(A.2) in \cite{Cremmer:1999du}, to
exhibit the connection with group theory. Having
claimed that the metric degrees of freedom take values in the vielbein
$SL(\tD\!-\!2)/SO(\tD\!-\!2)$, we expect that reducing the vielbein, we
have to decompose ($\R \cong SO(1,1)$ is the unique 1-dimensional
non-compact group, see appendix \ref{details})
$$
SL(\tD\!-\!2) \rightarrow SL(D \! - \! 2) \times SL(n) \times \R.
$$
Identifying $SL(D \! - \! 2)$ with the lower dimensional vielbein,
$SL(n)$ is represented in eq. (\ref{met}) by the appearance of the
weights $f_i^{\perp}$. Factors corresponding to the
remaining $\R$ factor have been split off. Commuting with $SL(n,\R)$
means that its representation vector is orthogonal to the
$f_i^{\perp}$, hence proportional to $s$. Tracelessness of the $\R$
generator fixes the ratio between the exponent of the factor in front of
the $D$ dimensional metric, and the one before the compact metric,
to $-(D-2)/n$. The overall values are fixed by the normalization of
the dilaton kinetic terms in the Lagrangian.  

With respect to the $\tilde{\gamma}^i_{\phantom{i}j}$ we note that,
after having set the vectors $\hat{\cA}_i=0$, they are part of a
vielbein on the compact space. The metric on the compact space is
given by 
$$
\dif s_n^2 = \exp(-\inp{f_i^{\perp}}{\phi})\delta_{ij}
\tilde{\gamma}^i_{\phantom{i}k} \tilde{\gamma}^j_{\phantom{j}l} \ \dif
z^k \otimes \dif z^l.$$ 
As in \cite{Lu:1995yn,Cremmer:1997ct, Cremmer:1999du} we also define
the inverse of $\tilde{\gamma}$ by $\gamma=\tilde{\gamma}^{-1}$. 

The Lagrangian in $\tD$ dimensions is 
\be 
{L}_{\tD} = R \df{\mathbf{1}} - \hlf \df{F}_{(p)} \wedge F_{(p)}
\ee
After reduction on the $n$-torus, the Lagrangian in $D$ dimensions
becomes  
\bea 
{L}_{D} & = &  R \df{\mathbf{1}} - \hlf \inp{\df{\dif \phi}}{\dif
  \phi}-\hlf e^{\inp{a}{\phi}} \df{F}_{(p)} \wedge F_{(p)} \non \\ 
 & &  -\hlf \sum_i e^{\inp{a_i}{\phi}} \df{F}_{(p-1)i} \wedge
F_{(p-1)i}-\hlf \sum_{i,j} e^{\inp{a_{ij}}{\phi}} \df{F}_{(p-2)ij}
\wedge F_{(p-2)ij}- \ldots \label{redlag}\\  
& &  -\hlf \sum_i e^{\inp{b_i}{\phi}} \df{\cF}^i_{(2)} \wedge
\cF^i_{(2)}-\hlf \sum_{i,j} e^{\inp{b_{ij}}{\phi}} \df{\cF}^i_{(1)j}
\wedge \cF^i_{(1)j} \non 
\eea
We have followed Cremmer et al. \cite{Cremmer:1999du} and defined the
following dilaton vectors in (\ref{redlag}) 
\be
{a} = -(p-1)s \qquad {a}_{i_1\ldots i_p} = \sum_{k=1}^p f_{i_k}-(p-1)s
\qquad {b}_i = -f_i \qquad {b}_{ij} = f_j - f_i 
\ee
In case one already has a dilaton and dilaton
couplings in the top dimension, one simply forms direct sums of the
dilaton vector before reduction, with the above dilaton vectors, and
couples these to $\tD$-dimensional dilatons and metric scalars in the
obvious way. 

Noting that
\be
\sum_{k=1}^n f_{k} = (\tD-2)s
\ee
we see
$$
a = -(\sum_{k=1}^n f_{k} - (\tD-p-1)s); \qquad a_i = -(\sum_{k \neq i}
f_{k} - (\tD-p-1)s); \qquad \textrm{ etc.}  
$$
These identities are a consequence of Poincar\'e duality in dimensionally
reduced theories: Reducing a form, its dilaton prefactor is the
inverse of the one found by reducing the
Poincar\'e dual form precisely over those dimensions that the form was
not reduced over \cite{Duff:1994an}. Kaluza-Klein vectors coming from
the metric obviously do not follow this pattern. 

The field strengths appearing in (\ref{redlag}) are defined as follows:
\be
F_{(q)i_1\ldots i_{p-q}} =
\gamma^{j_1}_{\phantom{j_1}i_1} \ldots
\gamma^{j_{p-q}}_{\phantom{j_{p-q}} i_{p-q}}
\hat{F}_{(q)j_1 \ldots j_{p-q}} \qquad \cF^i_{(2)} =
\tilde{\gamma}^i_{\phantom{i}j}\hat{\cF}^j_{(2)} \qquad \cF^i_{(1)j} =
\gamma^k_{\phantom{k}j} \hat{\cF}^i_{(1)k},  
\ee
with the hatted field strengths defined as
\bea
\hat{F}_{(p)} & = & \dif A_{(p-1)} - \dif A_{(p-2)i}\hat{\cA}^i_{(1)}
+ \hlf \dif A_{(p-3)ij}\hat{\cA}^i_{(1)} \hat{\cA}^j_{(1)} \non \\
& & -\frac{1}{6} \dif A_{(p-4)ijk} \hat{\cA}^i_{(1)} \hat{\cA}^j_{(1)}
\hat{\cA}^k_{(1)} \ldots \non \\
\hat{F}_{(p-1)i} & = & \dif A_{(p-2)i} + \dif
A_{(p-3)ij}\hat{\cA}^j_{(1)} + \hlf \dif A_{(p-4)ijk}\hat{\cA}^j_{(1)}
\hat{\cA}^k_{(1)} \non \\
& & +\frac{1}{6} \dif A_{(p-5)ijkl} \hat{\cA}^j_{(1)}
\hat{\cA}^k_{(1)} \hat{\cA}^l_{(1)} \ldots \\
& & \vdots \non \\
\hat{F}_{(1)i_1\ldots i_{p-1}} & = & \dif A_{(0)i_1\ldots i_{p-1}}\non
\\
\hat{\cF}^i_{(2)} & = & \dif \hat{\cA}^i_{(1)} \qquad
\hat{\cF}^i_{(1)j} = \dif \hat{\cA}^i_{(0)j} \non
\eea

The hatted field strengths are solutions to the identity
\be 
\dif \hat{F}_{(p-q)i_1\ldots i_{q}}= (-)^p \sum_j
\hat{F}_{(p-q-1)i_1\ldots i_{q}j} \hat{\cF}^{j}_{(2)} 
\ee
Using $\dif \tilde{\gamma}^i_{\phantom{i}j} = \hat{\cF}^i_{(1)j}$ and
$\dif \gamma^i_{\phantom{i}j} = -\cF^k_{(1)j} \gamma^i_{\phantom{i}k}$
one finds:  
\bea
\dif F_{(p-q)i_1\ldots i_{q}} & = &- \sum_{k} \cF_{(1)i_j}^{k}
\wedge F_{(p-q)i_1 \ldots k \ldots i_{(q)}}+ (-)^p\sum_{j}
F_{(p-q-1)i_1\ldots i_{q}j}\wedge \cF^{j}_{(2)}; \label{redeq1}\\ 
\dif\cF^i_{(2)} & = & \cF^i_{(1)j}\cF^j_{(2)}; \qquad \dif \cF^i_{(1)j}
= \cF^i_{(1)k}\cF^k_{(1)j}. \label{redeq2} 
\eea
These important identities are the Bianchi identities for the
dimensionally reduced theory. They form a vital clue to our discussion
in section \ref{oxi}. There we will use a different bookkeeping
for the indices. 

With the reduced Lagrangian given we may dualize a certain number of
fields, to reach the formulation of the theory we desire.  Note that
once a form is fully reduced, the length of its dilaton vector takes a
value independent of $D$ (but not of $\tD$!)
\be
| \sum_{k=1}^p f_{i_k} - (p-1) s |^2 = \frac{(p-1)(\tD-p-1)}{\tD-2}
\ee
Notice the symmetry under $p \leftrightarrow \tD - p$.
This result hints at the significance of $3$-forms in 11-dimensions,
and $4$-forms in 10 dimensions, as then a fully reduced form can play
the role of a root of a group lattice (having length squared
$2$). Other significant combinations are $2$ forms in 6 dimensions,
and $1$-forms in 5 dimensions, possibly giving rise to short roots of
groups. Indeed, all these cases are realized in the $E_8$, $E_7$,
$B_3$ and $G_2$ models, and give rise to a Freudenthal-like
construction of the groups as the composition of an $SL$ group with
some of its representations. All have supersymmetric extensions,
except for the $E_7$ theory. The latter can however be viewed as a
truncation of IIB supergravity.

\subsection{Projection to sublattices}\label{projsub} 

Reducing gravity from $\tD$ to 3 dimensions gives a
$SL(\tD-2)/SO(\tD-2)$ coset\cite{Cremmer:1998px}. So suppose that we
are given a coset $G/H$ parametrizing a scalar theory coupled to
gravity in 3 dimensions. If this can be oxidized, part of the coset
must be a reduced gravity theory, and hence we expect $SL(\tD-2)$ to be
contained in $G$, for some value(s) of $\tD$. The first step in
oxidation consist of identifying the $SL(\tD-2)$ sublattice (recall
that we are dealing with a regular subgroup) in the lattice spanned by
the dilaton vectors. Having found such a sublattice, we select a
sublattice of this sublattice corresponding to $SL(\tD-3)$. As, in a
sense, a $SL(\tD-3)$ group is geometrical, coming from the symmetry of
the torus, we can regard this as recovering the geometry. There are
$\tD-3$ positive roots of $SL(\tD-2)$ that are not roots of
$SL(\tD-3)$. We label them $g_i$, $i=1,\ldots \tD-3$, and order
them, $g_1$ being the highest root in the set, $g_2$ the highest except
for $g_1$ etc. From the $g_i$ we can build all the positive roots of
$SL(\tD-3)$: they have the form $g_i - g_j$, with $i < j$. Hence, at
least in 3 dimensions, we can identify $g_i = f_i$. We can then also
find a vector $t$, defined by $\sum g_i = (\tD -2) t$. Of course, in 3
dimensions $t=s$. 

The vectors $g_i$ and $(g_i - g_j)$ form the positive roots of the
$SL(\tD-2)$ lattice. We now wish to oxidize to $D$ dimensions, and
according to our earlier story, we have to decompose $G$ in an
$SL(D \! - \! 2)$ group plus complement. Without loss of generality,
we can assume that $SL(D \! - \! 2)$ has as simple roots $g_{\tD-3} $, and if
$D > 4$ the roots $g_{\tD-3-k}-g_{\tD-2-k}$, with $1 \leq k \leq D-4$. The
orthogonal complement to span($g_{\tD-3},g_{\tD-4}, \ldots$) then gives the
U-duality group in $D$ dimensions. We will give the results of the
projection.  

The projection of $g_i$ (where we assume that $i$ is not in the set
$\tD-3,\tD-4,\tD-5, \ldots$) we denote by $g_{P,i}$, the projection of $t$ by
$t_P$. They are given by 
\be
g_{P,i} = g_i - \frac{1}{D-2}\sum_{k=0}^{D-4} g_{\tD-3-k} \quad t_P = t -
\frac{1}{D-2} \sum_{k=0}^{D-4} g_{\tD-3-k} 
\ee
One easily verifies that
\be
\inp{g_{P,i}}{g_{\tD-3-k}} = 0 \qquad \inp{t_P}{g_{\tD-3-k}}=0 \quad 0 \leq k
\leq D-4
\ee
More crucial is that these vectors obey
\be
\inp{t_P}{t_P} = \frac{n}{(\tD-2)(D-2)}  \qquad \inp{t_P}{g_{P,i}} =
\frac{1}{D-2} \qquad \inp{g_{P,i}}{ g_{P,j}} = \delta_{ij} +
\frac{1}{D-2} 
\ee
Comparing these equations with (\ref{relations}) we see that the
$g_{P,i}$ form a set of $n$ independent vectors, and having precisely
the same innerproducts as the $f_i$ appropriate for $D$
dimensions. Hence the $n$ dimensional subspace spanned by the
$g_{P,i}$ and the $\R^n$ spanned by the $f_i$ can be rotated into
each other, such that the $g_{P,i}$ and $f_i$, and $t_P$ and $s$,
coincide. We can therefore switch to previous notation, and denote the
vectors again by $f_i$ and $s$.   

This demonstrates how to recover the lattice of dilaton vectors for
the higher dimensional theory from the coset in 3 dimensions. In the
next section, we will recover the full theory. 

\section{Increasing the number of dimensions: Oxidation}\label{oxi}

With the formalism of \cite{Lu:1995yn, Cremmer:1997ct,
  Cremmer:1999du}, reviewed in the previous subsection, one can
  construct, starting in the maximal dimension, the lower dimensional
  reduced theories. This section starts at the other end: Suppose we
  do have a certain scalar coset in 3 dimensions, how to reconstruct
  the higher dimensional theories?  

\subsection{Coset sigma models}

To come to grips with the full structure, we first study the relevant
coset sigma models.

Given a group $G$ with maximal compact subgroup $H$, there are 2
convenient forms for an action. Consider a group element $\cV \in
G$. We form the invariant ``metric'' $\cM = \cV^{\#} \cV$. The sigma
model Lagrangian can be written as: 
\bea
{L}_{G/H} & = & \frac{1}{4}e\ \textrm{tr}(\partial \cM^{-1}
\partial \cM) \label{maction}\\
& = & -e \ \textrm{tr}\left((\partial \cV)  \cV^{-1}\hlf(1+T)(\partial
\cV) \cV^{-1}\right) \label{taction}
\eea
where we have introduced the operator $T$, acting as $T(a) =
a^{\#}$. We use the superscript ${}^{\#}$ for ``generalized transpose'',
as in \cite{Cremmer:1997ct}. We have included a discussion in
our appendix \ref{Lie}. Because $(a^{\#})^{\#}=a$, the expression
$\hlf(1+T)$ represents a projection operator. The first
form of the action (\ref{maction}) features the metric $\cM$,
obviously constant under the action of $H$, while covariant under
$G$. We will focus on the second form of the action (\ref{taction}),
featuring $(\partial \cV) \cV^{-1}$ which is invariant under global
$G$ transformations, and transforms as a connection 1-form under
local $H$ transformations.  

The 1-form $(\partial \cV) \cV^{-1}$ is a tangent form on group
space, and hence it can be expressed in the group generators ($r$
denoting the (real) rank) 
\be \label{tform}
(\dif \cV) \cV^{-1} = \hlf \sum_{i=1}^r \dif {\phi}^i {H}^i +
\sum_{\alpha \in  \Delta^{+}} e^{\hlf{\inp{\alpha}{\phi}}}
F_{(1)\alpha} E_{\alpha}  
\ee
The second sum runs only over the set of positive roots $\Delta^+$,
because by virtue of the Iwasawa decomposition, we can always use an $H$
transformation to set terms with negative roots to zero. Henceforth,
we will work in this positive root gauge. One can then decompose
$\cV=\cV_H \cV_E$, with $\cV_H = \exp (\hlf \phi \cdot H)$ an
exponentiated element of the Cartan subalgebra, and $\cV_E$ generated
by the ladder operators $E_{\alpha}$ associated to the positive
roots. Then  
$$
(\dif \cV) \cV^{-1} = \dif(\cV_H^{\phantom{-}})\cV_H^{-1} +
\cV_H^{\phantom{-}} \dif(\cV_E^{\phantom{-}}) \cV_E^{-1} \cV_H^{-1} 
$$
which is the reason for the appearance of the exponential prefactors
in (\ref{tform}). 

Inserting (\ref{tform}) in the action (\ref{taction}), we obtain (with
the wedge product understood in the first term)
\be \label{cosaction}
{L}_{G/H} = - \inp{\df{\dif \phi}}{\dif \phi}-\hlf
\sum_{\alpha \in \Delta^+}e^{\inp{\alpha}{\phi}} \df{F}_{(1)\alpha}
\wedge F_{(1)\alpha} 
\ee

The $F_{(1)\alpha}$ are one forms. To get more detail on these, we
take the derivative 
\bd
\dif((\dif \cV) \cV^{-1}) = -(\dif \cV)\wedge(\dif \cV^{-1}) = ((\dif
\cV) \cV^{-1})\wedge((\dif \cV) \cV^{-1}) 
\ed
Substituting, one finds
\bea
\sum_{\alpha \in \Delta^+} \dif(e^{\hlf \inp{\alpha}{\phi}}
F_{(1)\alpha}) E_{\alpha} & = &    \frac{1}{8}\sum_{i,j} \dif \phi_i
\wedge \dif \phi_j [H_i, H_j] + \hlf \sum_{\alpha \in \Delta^{+}} e^{\hlf
  \inp{\alpha}{\phi}} \dif \phi_i \wedge F_{(1)\alpha}
    [H_i,E_{\alpha}] \non \\ 
& & + \hlf \sum_{\beta, \gamma \in \Delta^{+}}
    e^{\hlf\inp{\beta+\gamma}{\phi}} F_{(1)\beta} \wedge F_{(1)\gamma}
    [E_{\beta},E_\gamma] \label{dvv} 
\eea

Substituting the commutators, working out the differentials, and
checking per component (by multiplying each side with
$E_{-\alpha}$ for some $\alpha$, and taking the trace) one finds 
\be \label{scbi}
\dif F_{(1)\gamma} = \hlf \sum_{*} N_{\alpha,\beta} F_{(1)\alpha}
\wedge F_{(1)\beta} \qquad * = \left\{\ba{l} \alpha, \beta, \gamma \in
\Delta^+\\  
\alpha+ \beta = \gamma \ea \right.
\ee

The right hand side of this equation is symmetric in $\alpha$ and
$\beta$: interchanging the two gives a minus sign from the forms, and
another from $N_{\alpha,\beta}$. Because of the symmetry, each term
occurs 2 times; the factor of $\hlf$ cancels this.  

These Bianchi identities are an important clue. We want to arrive at
equations that do not explicitly mention $\cV$ (which is not an
invariant object). The Bianchi identities are an alternative way of
expressing that the right hand side of (\ref{tform}) can be
written as its left hand side, that is, as a tangent form on $G/H$;
they form a set of integrability conditions that ensure this. As such,
they encode group theoretic information, previously stored in
$\cV$. There is a field $F_{(1)\alpha}$ for every positive root $\alpha$,
the structure of (\ref{scbi}) is determined from the geometry of the
root lattice of $G$, and the structure constants $N_{\alpha,\beta}$
appear. Note that the positive roots of lowest height, the
simple roots, satisfy standard Bianchi identities.  

The structure becomes even nicer when also considering the equations
of motion for the scalar coset. There are two ways of obtaining these
from the action (\ref{cosaction}). The standard way would be to regard
the equations (\ref{scbi}) as Bianchi identities, solve for
$F_{(1)\alpha}$ in terms of potentials from these, and then derive the
equations of motion from (\ref{cosaction}) with the solutions for the
$F_{(1)\alpha}$ inserted. This is fairly tedious if $\textrm{dim }G$ is
large, and hides some of the nice covariance properties that will show
up in the following. 

Instead we regard the $F_{(1)\alpha}$ as independent fields, and enforce
(\ref{scbi}) by Lagrange multipliers (which will be
($D-2$)-forms). Therefore we add to the Lagrangian 
\be
{L}_{Bianchi} = \sum_{\alpha \in \Delta^+}\left(\dif
F_{(1)\alpha} - \sum_{\alpha=\beta+ \gamma} N_{\beta, \gamma}
F_{(1)\beta} \wedge F_{(1)\gamma} \right)\wedge A_{(D-2)-\alpha} 
\ee
The labels $-\alpha$ appearing on the $(D-2)$-forms will turn out
meaningful. 

Varying with respect to $A_{(D-2)-\alpha}$ will return the Bianchi
identities. Varying with respect to $F_{(1)\gamma}$ we find the
equations  
\be \label{defmin}
F_{(D-1)-\gamma} \equiv e^{\inp{\gamma}{\phi}}\df{F}_{(1)\gamma} =
\dif A_{(D-2)-\gamma}-\sum_{\beta-\alpha=-\gamma} N_{\beta,-\alpha}
F_{(1)\beta}  \wedge A_{(D-2)-\alpha} 
\ee
We defined $F_{(D-1)-\alpha}$, and used the identity (\ref{strcycl}) for the
structure constants. In the end, we are not interested in the dual potential
$A_{(D-2)-\alpha}$, but in the field strengths
$F_{(D-1)-\alpha}$. Taking the derivatives of these (note that in
general $\dif F_{(1)\beta} \neq 0$ !), one finds (see appendix
\ref{dereq} for details of the derivation)  
\be \label{sceqmo}
\dif F_{(D-1)-\gamma}= \sum_*  N_{\alpha,-\beta} F_{(1)\alpha} \wedge
F_{(D-1)-\beta}\qquad * =\left\{\ba{l} \alpha,\beta,\gamma \in
\Delta^+\\  \alpha-\beta=-\gamma \\
\ea \right. .
\ee
This should be compared to (\ref{scbi}). Notice the absence of a
factor $\hlf$. 

In terms of $F_{(D-1)-\alpha}$, the Lagrangian can be rewritten to
\be
{L}_{G/H} = - \inp{\df{\dif \phi}}{\dif \phi}-\hlf
\sum_{\alpha \in \Delta^+} F_{(D-1)-\alpha} \wedge F_{(1)\alpha},  
\ee
from which one finds the equation of motion for $\phi$
\be \label{cartaneqmo}
2 \dif(\df{\dif \phi}^i) = \sum_{\alpha \in \Delta^+} \alpha^i
F_{(D-1)-\alpha} \wedge F_{(1)\alpha} 
\ee
Note that with (\ref{cartaneqmo}), (\ref{scbi}) and (\ref{sceqmo}) we
have established a one-to-one relation between the conventional basis
of the Lie-algebra of $G$, and the equations relevant to the coset
sigma model. For every $E_{\alpha}$ where $\alpha$ is a positive root,
we have a Bianchi identity from (\ref{scbi}), when $\alpha$ is a
negative root we have an equation of motion from (\ref{sceqmo}), while
the Cartan subalgebra determines the equations for $\phi$. In
particular, the Cartan subalgebra only gives us $\textrm{rank }G$
equations, because the ``Bianchi identity'' for the potential for
$\phi$, $\dif^2 \phi = 0$ is trivial. This is a marked difference
between our philosophy, and the one from \cite{Cremmer:1998px}. For
the special case of the coset manifolds found from compactifying 11-d
supergravity the equations for the ``double'' of the fields as found
in \cite{Cremmer:1998px} coincide with our field equations.   

\subsection{Coupling to other fields}

Coupling to other forms $F_{(n)}$ is most easily done by adding
quadratic terms to the action. If the $F_{(n)}$ form a non-trivial
representation of U-duality, we need to contract on the internal
metric to render the action U-duality invariant. This takes the form 
\be \label{matteraction}
{L}_m = \frac{1}{2}\df{F}_{(n)} \wedge \cM F_{(n)}
\ee
We note that $\cM$ (or rather $\cV$) should be in an appropriate
representation. We will not discuss the possibility of Chern-Simons
modifications in this subsection, as these will appear naturally in
our discussion later.   

The equation of motion for $F_{(n)}$ then becomes
\be 
\dif (\cM \df{F}_{(n)})=0
\ee
We find it more convenient to substitute $M= \cV^{\#} \cV$ in this
equation, and rewrite it to 
\be \label{eqmo1}
\left(\dif + (\dif \cV \cV^{-1})^{\#} \right) \cV \df{F}_{(n)} = 0
\ee
The expression between the brackets is a covariant derivative,
ensuring that the equation of motion is covariant under the compact
part of the U-duality group, which is a \emph{local} symmetry. We
rewrite the standard Bianchi identity $\dif F_{(n)}=0$ to 
\be \label{bi1}
\left(\dif - (\dif \cV \cV^{-1})\right) \cV F_{(n)} =0
\ee 

If the fields $\df{F}_{(n)}$ and $F_{(n)}$ represent the same
degrees of freedom, we cannot allow them to transform
differently. Hence, only local transformations $O$ with $O^{\#}O=1$
are allowed, and this is the restriction to the compact subgroup that
was imposed already.  

Both $\cV F_{(n)}$ and $\cV \df{F}_{(n)}$ represent a full
$G$ multiplet of fields. The components of a multiplet can be
labelled by their weights, hence we decompose into components by
writing 
\be
\cV F_{(n)} \equiv \sum_{\lambda \in \Lambda}
e^{\hlf\inp{\lambda}{\phi}} F_{(n)\lambda},  
\ee
where the sum runs over the weights $\lambda$ on the weight lattice
$\Lambda$ of the representation.

One can work out the Bianchi identity (\ref{bi1}) by inserting  $(\dif
\cV) \cV^{-1}$ from (\ref{tform}), to obtain 
\be \label{bi2}
\dif F_{(n)\lambda'}= \sum_{*} N_{\alpha,\lambda} F_{(1)\alpha} \wedge
F_{(n)\lambda} \qquad * = \left\{\ba{l} \lambda, \lambda' \in
\Lambda\\ 
\alpha \in \Delta^+\\
\alpha+ \lambda = \lambda' \\
\ea \right.
\ee
The constants $N_{\alpha,\lambda}$ can be computed. We will not need
them explicitly; when finding expressions like (\ref{bi2}) in the
future, we will find that the constants are already determined in the
derivation. 

The equation of motion (\ref{eqmo1}) can be rewritten similarly. In
this case, the covariant derivative acts on the dual form,
which belongs to a different, conjugate $G$ representation. The
weights for the conjugate representation are the negatives of the
weights of the representation, so we write 
\be
\cV \df{F}_{(n)} \equiv \sum_{-\lambda \in \ov{\Lambda}}
e^{-\hlf\inp{\lambda}{\phi}} F_{(D-n)-\lambda},  
\ee
where we denote the weight lattice of the conjugate representation by
$\ov{\Lambda}$.  Again the reader should note that the forms of degree
$D-n$ are 
\emph{not} the duals to $F_{(n)\lambda}$. Rather, 
\be \label{id}
e^{-\hlf\inp{\lambda}{\phi}} \df{F}_{(D-n)-\lambda} =
e^{\hlf\inp{\lambda}{\phi}} F_{(n)\lambda}  
\ee 

The equation of motion becomes
\be \label{eqmo2}
\dif F_{(D-n)-\lambda'}= \sum_{*} N_{\alpha,-\lambda} F_{(1)\alpha}
\wedge F_{(D-n)-\lambda} \qquad * = \left\{\ba{l} \lambda,\lambda' \in
\Lambda \\ 
\alpha \in \Delta^+\\
\alpha-\lambda=- \lambda' \\
\ea \right.
\ee

It can happen that form and dual form transform in a self-conjugate
representation; in theories with self-dual tensors they \emph{must} be
in such a representation. In that case the
equation of motion (\ref{eqmo2}) and Bianchi identity (\ref{bi2}) are
essentially the same equation, and we can consistently impose
self-duality. 

The Lagrangian for coupled matter becomes
\be
{L}_m = - \frac{1}{2} \sum_{\lambda \in \Lambda}
F_{(D-n)-\lambda} \wedge F_{(n)\lambda} 
\ee
The sum over $\lambda$ indicates a sum over the weights of the
representation. Again, we get one Bianchi 
identity, and one equation of motion, labelled by $\lambda$
resp. $-\lambda$. For self-dual representations constraint
equation and Bianchi identity imply each other, but since then
$\lambda$ and $-\lambda$ belong to the same representation we
precisely get as many equations as weights. 

With extra matter, the equation of motion for the dilatonic scalars
(\ref{cartaneqmo}) is modified to 
\be \label{cartaneqmo2}
2 \dif(\df{\dif \phi}^i) = \sum_{\alpha \in \Delta^+} \alpha^i
F_{(D-1)-\alpha} \wedge F_{(1)\alpha} + \sum_{\lambda \in \Lambda} \lambda^i
F_{(D-n)-\lambda} \wedge F_{(n)\lambda}, 
\ee
while (\ref{sceqmo}) is not modified. Note that singlet
representations of U-duality do not couple to $\phi$ (as for these,
$\lambda=0$).   

For the non-dilatonic fields we have used roots and weights as
labels. In the action, combinations of forms always occur such that:
the degrees sum up to $D$; their labels sum up to zero. In equations
we find also that various terms have to have same degrees, but also
that their labels sum up to the same vector, which is either a root or
a weight. These properties are easily traced back to symmetries. The
rule on addition of forms is implied by Lorentz symmetry in
the non-compact directions. The rule for addition of the vector and
weight labels follows from the U-duality group. The theory is
invariant under $\cV \rightarrow \cV U$, with $U$ a constant element
of $G$. With the expansion of $\dif \cV \cV^{-1}$, most of these
symmetries became implicit. An exception is when $U$ is an element
obtained by exponentiating an element of the Cartan sub-algebra, then
we have 
\be
\phi \rightarrow \phi + \zeta; \qquad F_{(n)\xi} \rightarrow
e^{-\hlf\inp{\xi}{\zeta}}F_{(n)\xi}.  
\ee
regardless of whether $\xi$ is a weight or a root. Covariance of the
equations of motion and Bianchi identities therefore requires all the
labels of various terms to add up on both sides. The exponential
factors included in the definitions of fields labelled by negative
roots (\ref{defmin}) and conjugate weights (\ref{id}) ensure the
transformation behavior implied by their labels.

\subsection{Oxidizing}\label{inter}

We now know the relevant equations for a $G/H$ scalar
coset theory coupled to gravity in 3 dimensions. We shall not consider
the addition of other scalar fields\footnote{But note that this is not
  an irrelevant  possibility: Reduction of a generic theory
  with gravity and  matter results in a $SL(n)/SO(n)$ theory with
  additional scalars in representations of $SL(n)$. The semi-simple
  Lie groups discussed here require fine tuning of dilaton factors and
  Chern-Simons terms.}. The equations are then given by the 
Einstein equation, and the equations (\ref{scbi}), (\ref{sceqmo}) and
(\ref{cartaneqmo}). 

We will now (re)-construct the higher dimensional theories. In section
\ref{projsub} we recovered the lattice of the higher dimensional
theory, by choosing an $SL(D \! - \! 2)$-sublattice, and decomposing with
respect to this lattice. The group $G$ decomposes into $SL(D \! - \!
2) \times U_{D}$. The representations in the decomposition are at
least the adjoints of $SL(D \! - \! 2)$ and of $U_{D}$, but in general
there are more irreps. Because $SL(D \! - \! 2)$ is a level 1
subgroup, there is only one adjoint irrep of $SL(D \! - \! 2)$ and the
other irreps are antisymmetric tensors.   

We choose a basis of $D-3$ simple roots $\alpha_i$ for
the $SL(D \! - \! 2)$-lattice (which is a sublattice of the
$G$-lattice). We order the indices of the $\alpha_i$ along a direction
of the Dynkin diagram of $SL(D \! - \! 2)$ (which is a single chain of
nodes). We then have:  
$$
\inp{\alpha_i}{\alpha_i}=2, \qquad \inp{\alpha_i}{\alpha_{i+1}}=
\inp{\alpha_{i+1}}{\alpha_i}=-1, \qquad \inp{\alpha_i}{\alpha_j}=0
  \textrm{ otherwise}.
$$ 
The $D-3$ fundamental weights $\lambda_j$ are defined by restricting
to the space spanned by the $SL(D \! - \! 2)$ root lattice, and
demanding  
\be
2\frac{\inp{\alpha_i}{\lambda_j}}{\inp{\alpha_i}{\alpha_i}} =
\inp{\alpha_i}{\lambda_j} = \delta_{ij},
\ee
where we used that $SL(D \! - \! 2)$ is simply laced. The $\lambda_j$ are
highest weights for the antisymmetric $j$-tensor representation of
$SL(D \! - \! 2)$. The statement that the irreps of $SL(D \! - \! 2)$
other than the adjoint are forms, means that among their weights there
is a highest weight, which is either $0$ for a singlet, or $\lambda_j$
for some value of $j$. We associate a form with degree $j+1$ to the
representation with highest weight $\lambda_j$. The singlets transform
in the adjoint of $U_{D}$. For the semi-simple part of $U_{D}$ we
choose a basis of positive roots, for Abelian factors a positive
direction. To the positive roots of the semi-simple part of $U_D$ we
associate forms of degree $1$, to the negative roots forms of degree
$D-1$ (the Cartan generators, and non semi-simple parts correspond to
the dilatons). 

The equations for the oxidized theory are found as
follows. The roots of the $SL(D \! - \! 2)$ subgroup represent
the graviton; in the oxidized theory, it is encoded in
the Einstein equation. For the other roots $\alpha$, we associate to
$\alpha$ the field $F_{(n)\alpha'}$,
where $n$ corresponds to the degree defined previously, and
$\alpha'$ is the projection of the root $\alpha$ to the plane
orthogonal to the $SL(D \! - \! 2)$ lattice (it defines a
``representation vector'' for $U_D$; if $U_D$ is semisimple it is a
weight). There will be ${D-2} \choose {n-1}$ fields $F_{(n)\alpha'}$,
the dimension of the antisymmetric $n-1$-tensor irrep of $SL(D \! - \!
2)$. If $n \neq 1, D-1$, this number is bigger than 1, and these
multiple fields give equivalent descriptions of the same field. In the
oxidized theories, they correspond to multiple degrees of freedom
grouped in a single $SL(D \! - \! 2)$ tensor.  

The equations for the forms, and the non-dilatonic scalars then become
\be \label{oxieq}
\dif F_{(n)\alpha'} = \hlf \sum_* \eta_{l,\beta;m,\gamma} \
N_{\beta,\gamma}  F_{(l)\beta'} \wedge F_{(m)\gamma'}  \qquad * =
\left\{\ba{l} l+m=n+1\\  \alpha'+ \beta' = \gamma'
\ea \right. .
\ee
We sum over all combinations such that the sum of the degrees and the
labels of the right hand side match with degree and label on the
left hand side. The constants $\eta_{l,\beta;m,\gamma}$ are sign
factors: $N_{\alpha,\beta}$ is antisymmetric, but the combination of
the 2 forms is not necessarily, and the insertion of
$\eta_{l,\beta;m,\gamma}$ will prevent such terms from vanishing pairwise. We
explain how to compute $\eta_{l,\beta;m,\gamma}$ in the next
subsection.   

The multiple representatives of $F_{(n)\alpha'}$ in the end all
give the same equation, as a consequence of group theory. The
dilatonic equation of motion is (\ref{cartaneqmo2}), as
usual. The above recipe can be trivially applied to 3-dimensional
theories themselves, $SL(1)$ being the trivial group, leading to
singlets only.  

Note that, whereas the right hand sides of equations (\ref{scbi}),
(\ref{sceqmo}), (\ref{bi2}) and (\ref{eqmo2}) always involve at least
a one form appearing once, we have made no such restriction in
(\ref{oxieq}). These additional terms clearly go beyond the
coset structure defined by (\ref{scbi}), (\ref{sceqmo}), (\ref{bi2})
and (\ref{eqmo2}), and are known to appear in the higher dimensional
theories we wish to construct.  
 
Interpreting equation (\ref{oxieq}) as a Bianchi identity, one reason
for the appearance of terms with exclusively higher forms 
is due to the dimensional reduction procedure: Not only KK-scalars,
but also KK-vectors coming from the metric couple to the
various fields (compare with (\ref{redeq1})). Another possibility is
that a higher dimensional field may have a definition that includes a
``Chern-Simons'' term modifying its Bianchi identity. Such
modifications are often found in supergravity theories. In oxidation,
one can recover such terms as hallmarks of symmetries that are only
manifest in lower dimensions.   

If (\ref{oxieq}) is interpreted as an equation of motion, then an
action for this equation of motion must have
terms beyond those considered in (\ref{cosaction}) and
(\ref{matteraction}), of the Chern-Simons type. Varying a
Chern-Simons term consisting of one bare potential and two field
strengths, bilinear terms with field strengths of degree higher than
one will appear in the equations of motion. Of course Chern-Simons
terms are common in supergravity theories; from the point of
view of oxidation they are again signs of lower dimensional
symmetries. The oxidation procedure fixes the coefficient of
the Chern-Simons term to a structure constant of the algebra; this is
to be compared with the old observation that its value is
crucial to get the right symmetry in lower dimensions. 

\subsection{Sign factors}

For completeness, the sign factors
$\eta_{l,\beta;m,\gamma}$, introduced in
(\ref{oxieq}), have to be specified. Their computation is rather
technical; however, for many purposes it is not necessary to know the
explicit signs.

From our analysis of coset sigma models we easily deduce
\be
\eta_{1,\beta;1,\gamma} =1 \qquad \eta_{1,\beta;D-1,\gamma} =1 \qquad
\eta_{D-1,\beta;1,\gamma} = (-1)^D
\ee

To find the other sign factors, we have to deal with a
complication. To oxidize, we claim that one has to decompose $G$ into
$SL(D \! - \! 2) \times U_D$. This can be done at the level of the
lattices. We decompose the $G$ lattice into an $SL(D \! - \! 2)$ and a
$U_D$ lattice; for the $3$ dimensional theory we need a positive root
decomposition. Either decomposition is only specified up to
lattice automorphisms. A way to ensure that one theory is
directly related to the other without invoking lattice
automorphisms, proceeds as follows. First, choose a basis of
positive roots of $G$, and draw the corresponding Dynkin diagram. Then
extend the diagram with the lowest root of the $G$ lattice (the
next section will feature more on extended Dynkin diagrams). Now
choose an $SL(D \! - \! 2)$ subdiagram of the extended $G$ diagram, such
that one of the ends of the $SL(D \! - \! 2)$ chain coincides with the
extended node; this guarantees that the positive roots of $U_D$ are
all made up of positive roots of $G$. To relate the theories by
dimensional reduction, one must choose the right orientation of
the $SL(D \! - \! 2)$ diagram; this is done by attaching the highest
index to the extended node, and labelling downwards along the chain.  

Consider a positive root $\alpha$ of $G$. To find its
representative in the oxidized theory, decompose with respect to
the $SL(D \! - \! 2)$ lattice. Let $\beta$ be the component parallel
to the space spanned by the $SL(D \! - \! 2)$ lattice; $\beta$ must be
a weight of an antisymmetric tensor representation of $SL(D \! - \!
2)$, say the $k$-form. Calling the complement of $\beta$ $\alpha'$,
the oxidation recipe tells us that the fields in the 3-dimensional and
the $D$-dimensional theory are  
$$\alpha = \alpha' + \beta  \qquad \Rightarrow  \qquad F_{(1)\alpha}
\rightarrow F_{(k+1)\alpha'}$$ 
One can do the same for the negative root $-\alpha$; decomposing the
weight $-\beta$ of the $D-2-k$ antisymmetric tensor irrep is
found. This leads to the representatives
$$-\alpha = -\alpha' - \beta \qquad \Rightarrow \qquad F_{(2)-\alpha}
\rightarrow F_{(D-k-1)-\alpha'}$$ 

Now suppose we have a term in the equations for the oxidized theory: 
\be
\dif F_{(n)\alpha'} = \ldots + \eta_{l,\beta;m,\gamma} \
N_{\beta,\gamma}  F_{(l)\beta'} \wedge F_{(m)\gamma'} + \ldots
\ee
We reduce straightforwardly on a rectangular torus (a non-rectangular
torus would only result in cluttering the equations with extra terms
due to KK-fields from the metric). To do so, we replace each field
$F_{(n)\alpha'}$ by either $F_{(1)\alpha}\dif z_{i_{n-1}} \ldots \dif
z_{i_1}$ or $F_{(2)\alpha}\dif z_{i_{n-2}} \ldots \dif z_{i_1}$,
whichever one is appropriate. The addition of differential forms $\dif
z_i$ leads to sign ambiguities in the reduced forms. As a matter of
fact, we can choose any convention as long as we do so
consistently. We will choose to order the indices on the $\dif
z_{i_j}$ in decreasing order, i.e. $i_j > i_k$ if $j > k$. Of course
the products of $\dif z_{i_j}$'s have to match on both sides.

The above example becomes then ($l',m', n'$ are $1$ or $2$, whichever one
is appropriate)
\be
\ba{l}
\dif F_{(n')\alpha} \dif z_{n-n'} \ldots \dif z_1 = \\
  \ldots + \eta_{l,\beta;m,\gamma} \
N_{\beta,\gamma}  F_{(l')\beta}\dif z_{n-n'} \ldots \dif z_{m-m'+1}
\wedge F_{(m')\gamma}\dif z_{m-m'} \ldots \dif z_1 + \ldots
\ea
\ee 
Reordering, and comparing with (\ref{sceqmo}) or (\ref{scbi}), one
finds (using that $l'$ is either $1$ or $2$)
\be \label{eta}
\eta_{l,\beta;m,\gamma}= (-1)^{(l-l')m'+l'+1}   
\ee
Note that running the same argument with $F_{(l)\beta'}$ and
$F_{(m)\gamma'}$ interchanged leads to
\be
\eta_{m,\gamma;l,\beta}= (-1)^{(m-m')l+m'+1}   
\ee
which is not the same as (\ref{eta}) with $l$ and $m$, $l'$ and $m'$
interchanged. Using that $l'm'+ l'+ m'$ is odd
(at most one of $l'$ or $m'$ is even), one easily shows that 
\be
\eta_{l,\beta;m,\gamma} = (-)^{lm+1}\eta_{m,\gamma;l,\beta},
\ee 
which is a necessary condition to prevent
terms from (\ref{oxieq}) from vanishing pairwise.

\subsection{Comparison to dimensional reduction}\label{proof}

The above oxidation recipe is simple and elegant, but still has to
be compared to what is obtained by dimensional reduction. To do so, we
construct theories from a $G$ lattice in $D+1$ and in $D$ dimensions,
and compare them with each other. Applying dimensional reduction to
the $D+1$ dimensional theory should give the $D$ dimensional theory. To
prove full equivalence, one can then pick any theory in the chain of
theories related by reduction and oxidation, and appeal to induction.  

According to the recipe, we have to decompose $G$ into $SL(D-1)$ and
$SL(D \! - \! 2)$ subgroups respectively. To relate the two, we embed
$SL(D \! - \! 2)$ in $SL(D-1)$, and compare their representations. The
adjoint of $SL(D-1)$ corresponds to the $(D+1)$ dimensional
graviton; it decomposes into adjoint, fundamental,
anti-fundamental and a singlet of $SL(D \! - \! 2)$. The adjoint is the
$D$-dimensional graviton; the fundamental and anti-fundamental
representations give equation of motion and Bianchi identity for a
single KK-vector; and the singlet is a dilatonic scalar, corresponding
to 1 equation. An $n$-form for $SL(D-1)$ decomposes into an $n$-form
and an $n-1$ form of $SL(D \! - \! 2)$. These decompositions are
clearly compatible with the dimensional reduction procedure.   

Next we compute the relation between the $(D+1)$ dimensional dilaton
factors and the $D$ dimensional ones (this partly doubles the
discussion in section \ref{projsub}, but we think this is
instructive). We will use that the lengths of the weights for the
antisymmetric $k$-tensor representation of $SL(m)$ are given by 
\be
\sqrt{\frac{(m-k)k}{m}}.
\ee
The $D$-dimensional KK-vector coming from the $(D+1)$-dimensional
metric is represented by roots of $SL(D\!-\!1)$ that are no longer a root
of $SL(D \! - \! 2)$. Rather, they are represented by the fundamental
irrep of $SL(D \! - \! 2)$. Knowing the lengths of the roots and
weights involved, the dilaton coefficient can be computed by using
Pythagoras' theorem:  
\be
\left(2 - \frac{D-3}{D-2}\right)^{\hlf} =
\frac{D-1}{\sqrt{(D-1)(D \! - \! 2)}}=x_2 
\ee
The computation for the forms proceeds similarly. For the $n$-form of
$SL(D)$, one finds
\be
\left(\frac{n(D-1-n)}{D-1} -\frac{n(D-2-n)}{D-2}\right)^{\hlf} =
\frac{n}{\sqrt{(D-1)(D-2)}}=x_{n+1} 
\ee
The $(n-1)$ form of $SL(D\!-\!2)$ gives
\be
\left(\frac{n(D-1-n)}{D-1} - \frac{(n-1)(D-1-n)}{D-2}\right)^{\hlf}
=\frac{D-n-1}{\sqrt{(D-1)(D-2)}}=x_{n} 
\ee
That $x_{n+1} + x_n = x_2$ is no coincidence, but a simple consequence
of the relations between the relevant weight lattices.
These are absolute values. Tracelessness of the algebra generators of
the $n$-tensor representation of $SL(D\!-\!1)$ requires the coeffients
$x_n$ and $x_{n+1}$ to have opposite signs. An explicit computation
reveals that $x_2$ has the same sign as $x_{n+1}$. The overall sign can be
absorbed in a redefinition of the dilatons, so we choose a
(positive) sign for $x_{n}$. 

In summary, the argument for the dilaton factor may be defined recursively
as  
\be
\alpha_{D} = (\alpha_{D+1}, x) \qquad x = \left\{\ba{ll} -x_{n+1} &
\textrm{for }F_{(n+1)} \rightarrow F_{(n+1)} \\
  x_{n} & \textrm{for }F_{(n+1)} \rightarrow F_{(n)} \\
 -x_2 & \textrm{for }{\cal F}_{(2)} \\ \ea \right. ,
\ee
demonstrating complete agreement with \cite{Lu:1995yn} (up to a factor
of $\sqrt{2}$ that we absorbed in $\phi$). Hence, upon dimensional
reduction, the labels referring to the U-duality grow as claimed. 

This is also sufficient to guarantee that the action has the right
form, with the right coefficients, and that the Einstein equation and
the equation for the dilatons take the right form. Complete agreement
requires that the field strengths for the forms and the axions are defined in
the same way: This is checked by comparing the Bianchi identities in
both formalisms.   

We first check that oxidation gives the right answer for the
fields coming from the reduction of the gravity sector. Pure
gravity is obtained by oxidizing the $SL(D \! - \! 2)/SO(D \! - \! 2)$-coset 
\cite{Cremmer:1998px}. Assuming we are not in the maximal dimension,
we need the following decomposition (see appendix \ref{coset},
$d+d'=n$) 
\be
\ba{rcl}
SL(n,\R) & \rightarrow & SL(d,\R) \times SL(d', \R) \times \R \\
\mathbf{(n^2-1)} & \rightarrow & \mathbf{(d^2\!-\!1,1)^0 \oplus
  (1,d'^2\!-\!1)^0 \oplus (1,1)^0  \oplus (d,d')^n \oplus (\ov{d},
  \ov{d'})^{-n}} \ea 
\ee
Setting $d$ to $D-2$, we find the fields for the $SL(d')/SO(d') \times
\R$ coset, and $d'$ vectors transforming in the fundamental
irrep of $SL(d')$. The $SL(d')$ adjoint gives us $d'-1$
dilatons we group in $\phi$, and there is an additional $\phi_{\R}$
for the $\R$ factor. The $\hlf d'(d'-1)$ positive roots of $SL(d')$
give rise to axion field strengths $F_{(1)\alpha,0}$, and the same
number of negative roots result in forms $F_{(D-1)-\alpha,0}$. The
label $\pm\alpha$ corresponds to a positive/negative root of $SL(d')$,
while we have added a third label for the charge under $\R$. The
fundamental irrep $\mathbf{d'}$ gives rise to $d'$ vector
field strengths $F_{(2)\lambda, n}$, with $\lambda$ a weight of the
irrep, while there are $d'$ forms $F_{(D-2)-\lambda,-n}$ from the
$\mathbf{\ov{d'}}$ irrep. 

Our recipe leads to the following Bianchi identities for the
axions and the form fields
\be
\ba{rcll}
\dif F_{(1)\gamma,0} & = & \hlf \sum_{*} N_{\alpha,\beta}
F_{(1)\alpha,0} \wedge F_{(1)\beta,0} &  * = \left\{\ba{l} \alpha,
\beta, \gamma \in \Delta^+\\ 
\alpha+ \beta = \gamma
\ea \right. \\ 
\dif F_{(2)\lambda',n} & = & \sum_{*} N_{\alpha,\lambda} F_{(1)\alpha,0}
\wedge F_{(2)\lambda,n} & * = \left\{\ba{l} \alpha \in
\Delta^+\\ 
\alpha+ \lambda = \lambda'
\ea \right. 
\ea
\label{abi}
\ee
Setting $N_{\alpha,\lambda}=1$, the equations (\ref{abi}) agree with
(\ref{redeq2}), in any dimension. 
 
For the form fields we explicitly compare the $D+1$ dimensional theory
with the $D$ dimensional one. An $n$-form of $SL(D-1)$ decomposes to
an $n$ form and an $n-1$ form of $SL(D \! - \! 2)$, both charged
under the dilaton appearing when reducing from $D+1$ to
$D$-dimensions. Their charges have opposite sign, and tracelessness of
the $n$ tensor representation of $SL(D-1)$ means that the charges of
the $n$-tensor resp. the $n-1$ tensor of $SL(D \! - \! 2)$ are $-n$ and
$D-1-n$. We also need that the axions are not charged under the
new dilaton. Then, for the $n$-tensor of
$SL(D \! - \! 2)$, our formalism predicts: 
\bea
\dif F_{(n+1)\lambda_n,-n} & = & \sum
N_{(\alpha,0),(\lambda_n,-n)}F_{(1)\alpha,0} \wedge
F_{(n+1),\lambda_n,-n} + \nonumber \\
& & \sum N_{(\lambda_{n},D-1-n),(0,D-1)} 
F_{(n)\lambda_{n}, D-1-n} \wedge F_{(2)0,D-1} \label{abi2}
\eea
We denoted the label of the $n$-tensor of $SL(D-1)$ by $\lambda_n$
and added the charges as a second label. The structure of
(\ref{abi2}) is identical to (\ref{redeq1}); closer inspection shows
that detailed matching can be done. The reader can convince himself that
addition of Chern-Simons terms modifies the equations in the way our
formalism predicts, in the formalism of \cite{Lu:1995yn} these have to
be added by hand.

\section{Extended Dynkin diagrams}\label{diag}

It has been known for long that aspects of the supergravity
  theory can be encoded in Dynkin diagrams. Extended Dynkin
  diagrams\footnote{These are also known as the Dynkin diagrams of the
  affine untwisted  Kac-Moody algebra's. They are however useful in
  many contexts where  a relation with Kac-Moody algebra's is not
  obvious.} encode even more information about the theory. This can be
  traced back to the original reason for their introduction, in
  \cite{Dynkin:um}, as a tool for the classification
  of regular subalgebra's of a given semi-simple Lie algebra. These
  diagrams have appeared previously in the supergravity literature
  (e.g. in \cite{Julia:1980gr, Julia:1982gx}); here we focus on
  applications to theories with dimension $\geq 3$. Appendix
  \ref{coset}, figure \ref{dynkfig} lists all extended Dynkin diagrams.

\subsection{Representation content of a theory}

We first recall some definitions. Given the root lattice for a
simple Lie group, one can choose a basis of linearly independent
simple roots, the simple roots being defined by the property that the
difference of two simple roots is not a root. The geometrical
relations between these basisvectors can be conveniently encoded in a
Dynkin diagram. 

Dropping the requirement of linear independence, the basis can be
extended by adding the lowest root of the root lattice to the
basis. Though no longer independent, all elements of such a basis
still have the property that the difference of two of them is not a
root. As a consequence, any proper subset of the vectors of the
extended basis defines a set of simple roots for some lattice, and the
lattice in question is a sublattice of the original group
lattice. The extended basis can also be encoded in a diagram, the
extended Dynkin diagram, and the process of dropping some basis
elements corresponds to erasing some nodes of the diagram. Repetition
of the procedure (extension of the diagram, choosing a subdiagram)
allows one to find all the regular sublattices \cite{Dynkin:um}. 

Applied to our problem, the sublattices we are interested in are the
$SL(n)$ sublattices, realized as chains in the extended Dynkin
diagram. In the generic case, we obtain a disconnected diagram when
erasing nodes such that we obtain a $SL$-chain. The remaining diagram,
with the chain taken out, corresponds to the lattice of a (regular)
subgroup of the U-duality group in that dimension. If the extended
node is part of the $SL(n)$-chain, the complementary diagram precisely
gives the lattice of the semi-simple part of the U-duality group. If
the final (not extended) diagram has less nodes than the original (not
extended) diagram, the difference in nodes corresponds to $\R$ factors
in the U-duality group. It has been known for a long time
that ``group disintegration'' should start at the end of the Dynkin
diagram where one attaches the affine vertex (In \cite{Cremmer:1999du}
this is verified for all split simple Lie-groups, for the
$E_{8(8)}$-series of maximal supergravity the observation is much
older \cite{Julia:1982gx}). The above argument demonstrates why
this observation is correct.

As an immediate corollary, we notice that for every coset theory $G/H$
in 3 dimensions with split $G$, the maximal oxidation dimension can be
found by taking the extended diagram for $G$, looking for the largest
chain of nodes representing long roots, and counting the number of nodes of
this chain. Adding 3 to this number, we find the maximal oxidation
dimension. Comparing with the list in the summary section of
\cite{Cremmer:1999du}, we see that this simple rule covers all cases.

Another important point is that if there are multiple ways of
realizing the diagram of a sublattice, the various possibilities
correspond to subgroups that are conjugate to each other, \emph{except
  for a finite number of exceptions}, listed in Dynkins paper
\cite{Dynkin:um}. These play a central role in the next subsection. 

Before arriving at the final diagram, we have to erase a certain
node. We know the geometrical relations of this node to the final
sublattice, they were encoded in the diagram before erasing the
node. This gives us information on the representation content of the
theory. The Dynkin diagram encodes the entries of the Cartan matrix,
which is given by 
\be
N_{ij} = 2\frac{\inp{\alpha_i}{\alpha_j}}{\inp{\alpha_i}{\alpha_i}}
\ee
We are interested in $SL$ groups of long roots, for which we have
$\inp{\alpha_i}{\alpha_i}=2$. Hence
$N_{ij}=\inp{\alpha_i}{\alpha_j}=-1$, if the $j^{th}$ node was linked
to the $i^{th}$ one. This implies that the node that was
erased corresponds to a lowest weight for a representation, and the
node it was connected to reveals which representation: If the erased
node connected to the $k^{th}$ node of the $SL(n)$ chain (which has
$n-1$ nodes), the relevant representation is the $(n-k)$-form (see
also \cite{Julia:1997cy}). Of
course one can enumerate the nodes in the $SL(n)$ chain in 2 ways, and
one finds also a $k$-form (provided $2k\neq n$). But this should be
expected: The adjoint is a self-conjugate irrep, and should
decompose in selfconjugate representations. Hence for each $k$-form
one should also find the conjugate $(n-k)$-form, the exception being
$n/2$-forms for $n$ even, which are selfconjugate. In principle also
the representation of the U-duality group can be read off 
this way, but here we can also have $N_{ij}=-2,-3$ for non-simply
laced groups. Also, because the U-duality group is not an $SL$-group
in general, additional knowledge of the representations of the
U-duality group is required. 

\begin{figure}[ht]
\begin{center}
\includegraphics[width=10cm]{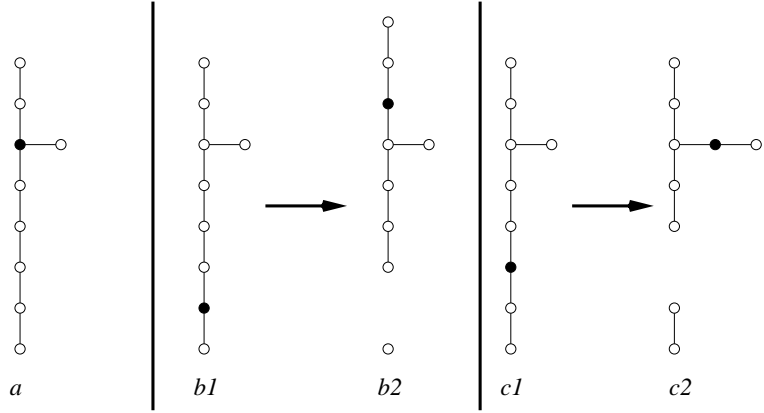}
\caption{Manipulations of Dynkin diagrams} \label{Dynkfig}
\end{center}
\end{figure}

As an example we depicted in figure \ref{Dynkfig} the manipulations
 that lead to the $SL(6) \times SL(2) \times SL(3)$ subgroup of
 $E_8$. This is relevant for the toroidal compactification of maximal
 supergravity to $6+2=8$ dimensions, where indeed the U-duality group
 is $SL(2) \times SL(3)$. The relevant Dynkin diagram can be obtained
 in various  ways. The simplest is by erasing the black node in the diagram
 denoted by $a$. This node is connected to the ends of the $A_1$,
 $A_2$ and  $A_5$ diagrams, hence we expect vectors
 (from the end of $A_5$), transforming as a doublet under $SL(2)$ and
 a triplet under $SL(3)$. Another way of arriving at the same diagram,
 depicted in $b1,b2$, proceeds by first erasing a node to obtain $A_1
 \oplus E_7$, then extending $E_7$ and erasing the black node in
 $b2$. This node is connected to the second node of the
 $A_5$ chain, and hence corresponds to 2-tensors. They are singlets
 under $SL(2)$ (not connected), and form a triplet under $SL(3)$. The third
 way of getting to the final diagram proceeds via $A_2 \oplus E_6$,
 and from this we learn that there are 3-tensors, in a doublet under
 $SL(2)$ and singlet under $SL(3)$. 

Instead of giving more examples, we encourage the reader to play with
 the extended Dynkin diagrams to recover the matter content of the
 relevant theories. For comparison, the relevant representations are
 listed in appendix \ref{coset}. 

\subsection{Dualities}

A interesting phenomenon is the appearance of diagrams with
vertices. At such a vertex, an embedded chain, corresponding to an
$SL(n)$ group can turn in different ways. The corresponding $SL(n)$
groups are non-conjugate: They are listed in \cite{Dynkin:um}.

\begin{figure}[ht]
\begin{center}
\includegraphics[width=10cm]{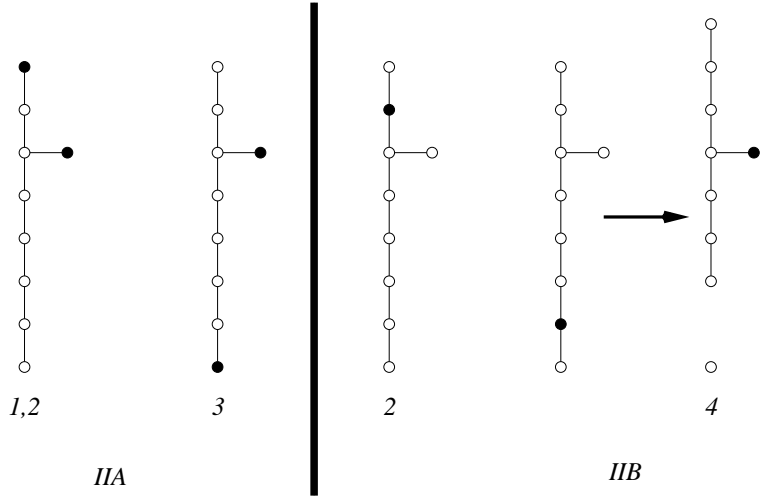}
\caption{The extended Dynkin diagram of $E_8$: IIA vs. IIB}
\end{center}
\end{figure}

A very prominent example can be found inside the $E_8$ diagram. There
are two different, and inequivalent ways of embedding an $SL(8)$ chain
in this diagram. In one case we obtain a 10 dimensional theory with
U-duality group $\R$. The various ways of embedding the chain inside
the ``long end'' of the $E_8$ diagram reveal a vector, a 2-tensor, and
a 3-tensor, and hence this theory should be identified with IIA
supergravity. A second possibility is to insert the $A_7$ chain in the
``short'' end. Its complement is then an $SL(2)$, and the erased node
reveals a doublet of 2-tensors. A second way to reach $SL(8) \times
SL(2)$ proceeds by first breaking $E_8$ to $E_7 \times SL(2)$, and
subsequently breaking $E_7$ to $SL(8)$. This path to the duality group
reveals a 4-tensor, which is a singlet under $SL(2)$. Of course this
second theory is IIB supergravity. The $SL(8)$ of IIB can not be
embedded in the $SL(9)$ of 11 dimensional supergravity, but smaller
$SL(n)$ subgroups can be embedded in either the $SL(8)$ of IIB-theory,
or the $SL(9)$ of 11 dimensional supergravity. 

The other examples of diagrams with vertices are the $B_n$ and $D_n$
series, and $E_6$ and $E_7$.

The $D_n$-diagrams with their two vertices are particularly
interesting. The $D_8$ diagram corresponds to type I supergravity,
that also occurs in the heterotic string. Because of the forks there
are two different ways of embedding an $A_7$ chain. Comparing the two
options, one sees that the two different ways correspond to exchanging
moduli coming from the metric with moduli that come from a 2-tensor:
Again this represents T-duality in this context. Also interesting
are the two inequivalent embeddings\footnote{In Dynkins paper
  \cite{Dynkin:um} the inequivalent $SL(4)$'s are denoted as $A_3$ and
  $D_3$, respectively.} of $SL(4)$. One of them corresponds to
compactifying the bosonic sector of 10 dimensional type I supergravity
to 6 dimensions, the other is a theory of gravity with only self-dual
and anti-selfdual 2-tensors, and corresponds to the bosonic sector of
$(2,0)$ gravity in 6 dimensions with tensormultiplets. The theories
can be related in 5 dimensions, as they can be regarded as truncations
of the compactification of IIA and IIB string theory on $K3$, and adding an
additional circle makes T-duality possible. The reader should
compare the number of vectors found by group theory with the
discussion in \cite{Witten:1995ex}. 

Intriguingly, these well known dualities seem to be present in all the
$D_n$ theories (at least as far as the massless field are concerned),
in particular in the $D_{24}$ theory, which corresponds to the
massless sector of the bosonic string. Also this gravity theory has a
branch in 6-d, with 21 self-dual and 21 anti-selfdual tensors
transforming in a single vector irrep of $SO(21,21)$. Though it is not
obvious that this is possible, it would be interesting to attempt to
construct a string theory realizing these massless fields. Doing so
might provide clues on issues such as little string theories, and a
possible role for the bosonic string in string dualities. Let us also
point out the $D_4$ extended Dynkin diagram, the only diagram with a
four-vertex, yielding a highly symmetric spectrum upon
compactification. It gives rise to a 6 dimensional theory that, upon
compactification on a circle exhibits a ``T-triality'' rather than
T-duality.  

The $B_n$ chains have similar symmetries as the $D_n$-series. Again
there is the possibility of different ways of embedding a ``long''
chain, and an additional 6 dimensional branch. The difference between
the $D_n$ and $B_n$ model in their maximal dimension is the presence
of an extra vector. Apart from the modification of the action, this
leads to a modification of the equation for the 2-form field
strength: It becomes
\be \label{anom}
\dif F_{(3)} = -\hlf F_{(2)} \wedge F_{(2)}
\ee
Note that for $n > 4$ ($D > 6$) this is usually interpreted as a Bianchi
identity, while for $n < 4$ ($D < 6$) it is an equation of motion. 
This can be seen by using group theory. Let $e_i$ be
unit vectors on the $\R^n$, then one can pick the following positive
roots for $B_n$: $e_i-e_j$, these will be the
$SL(n)$ in the maximal dimension; $e_i + e_j$, these correspond to the
antisymmetric tensor in the top dimension; and $e_i$, these form the
vector in the top dimension. One then immediately sees that one has to
find an equation of the form of (\ref{anom}). The
identity (\ref{anom}) is well known in string theory in a version
where the r.h.s. is replaced by $\textrm{Tr}(F_{(2)} \wedge F_{(2)})$;
it plays a crucial role in anomaly cancellation \cite{Green:sg} (A
second first 
Pontryagin class associated to the Riemann tensor does not turn up
at the level of classical supergravity, but is a higher derivative
effect). Our derivation here has nothing to do with supersymmetry or
string theory, but produces the term from the group structure. A
similar identity occurs when enlarging the $D_n=SO(n,n)$ or $B_n=
SO(n+1,n)$ groups to $SO(n+r,n)$; this also implies vectors in the
maximal dimension, and modification of the 2-form Bianchi identity, as
we will explain in \cite{Arjannew, HJK}.  

Although it is a bit outside the scope of the rest of this subsection, that
focuses on T-duality, note that the fork in $B_n$ and $D_n$
diagrams with $n \geq 2$ leads to a separate $SL(2,\R)$ factor in 4
dimensions, commonly associated to S-duality of gauge theories. Also note
that all of the $B_n$ and $D_n$ theories oxidize in their maximal
dimension to a theory with antisymmetric 2-tensors, and hence,
strings. Together with the fact that all these theories have a
6-dimensional branch, this stresses once more various known relations
between strings and gauge theories.
 
The $E_7$ theory is most easily interpreted as a truncation of
IIB-theory, by decomposing $E_8 \rightarrow E_7 \times SL(2)$, and
truncating to the sector that contains only the singlets under
$SL(2)$. The two inequivalent ways of embedding an $A_6$ chain lead to
a branch in $d=8$. This can also be seen from the viewpoint of $E_7$
theory as a truncation of a maximally supersymmetric theory. This
would have U-duality group $SL(2) \times SL(3)$, with the original IIB
$SL(2)$ embedded in $SL(3)$. Of course it is also possible to truncate
by the other $SL(2)$, that arises because the 3-form of 11 dimensional
supergravity reduced over a 3-torus forms a complex combination with a
dilaton, which transform under $SL(2)$. 

The extended diagram of $E_6$ is
highly symmetric. Viewing $E_6$ theory as a truncation of $E_8$
theory, comparison of the branches in the diagram shows a relation with
``M-theory T-duality''; a version of T-duality acting on 2 directions
of the compactification torus simultaneously. 

\section{Discrete symmetries: Magic triangles}\label{triang}

As a direct application of our results, we point out that the
discrete symmetries pictured in so-called magic triangles, can be
easily understood in the developed formalism. 

\subsection{Magic triangle}

In \cite{Cremmer:1999du}, a ``magic triangle'' appears. This is a table
of the U-duality groups appearing in the oxidation of $E_n$ cosets from
3 dimensions. Its content is given in table \ref{mt}. 

\begin{table}[ht]
\begin{center}
\begin{tabular}{|c|ccccccccc}
\cline{1-2}
11 & \mc{1}{|c|}{$\phantom{\Big(}\{ e \}\phantom{\Big)}$} & & & & & &
& &  \\
\cline{1-3}
10 & \mc{1}{|c|}{$\phantom{\Big(}{\R \atop A_1} \phantom{\Big)}$} &
\mc{1}{c|}{$\{ e \}$} & & & & & & & \\
\cline{1-3}
 9 & \mc{1}{|c|}{$\R \! \times \! A_1$} &
 \mc{1}{c|}{$\phantom{\Big(}\R\phantom{\Big)}$} &  & & & & & & \\
\cline{1-4}
 8 & \mc{1}{|c|}{$A_1 \! \times \! A_2$} &
 \mc{1}{c|}{$\phantom{\Big(}{{\R \times A_1} \atop
     A_2}\phantom{\Big)}$} & \mc{1}{c|}{$A_1$}  & & & & & & \\ 
\cline{1-6}
 7 & \mc{1}{|c|}{$\phantom{\Big(}A_4\phantom{\Big)}$} & \mc{1}{c|}{$\R
   \! \times \! A_2$} & \mc{1}{c|}{$\R \! \times \! A_1$} &
 \mc{1}{c|}{$\R$} & \mc{1}{c|}{$\{ e \}$} & & & & \\
\cline{1-6}
 6 & \mc{1}{|c|}{$D_5$} & \mc{1}{c|}{$A_1 \! \times \! A_3$} &
 \mc{1}{c|}{$\R \! \times \! A_1^2$} &
 \mc{1}{c|}{$\phantom{\Big(}{\R^2 \atop A_1^2}\phantom{\Big)}$} &
 \mc{1}{c|}{$\R$} & & & & \\
\cline{1-7}
 5 & \mc{1}{|c|}{$\phantom{\Big(}E_6\phantom{\Big)}$} &
 \mc{1}{c|}{$A_5$} & \mc{1}{c|}{$A_2^2$} & \mc{1}{c|}{$\R \! \times \!
   A_1^2$} & \mc{1}{c|}{$\R \! \times \! A_1$} & \mc{1}{c|}{$A_1$} & &
 & \\
\cline{1-9}
 4 & \mc{1}{|c|}{$E_7$} & \mc{1}{c|}{$D_6$} & \mc{1}{c|}{$A_5$} &
 \mc{1}{c|}{$A_1 \! \times \! A_3$} & \mc{1}{c|}{$\R \! \times \!
   A_2$} & \mc{1}{c|}{$\phantom{\Big(}{{\R \times A_1} \atop
     A_2}\phantom{\Big)}$} & \mc{1}{c|}{$\R$} & \mc{1}{c|}{$\{ e \}$}
 &  \\
\cline{1-10}
 3 & \mc{1}{|c|}{$E_8$} & \mc{1}{c|}{$E_7$} & \mc{1}{c|}{$E_6$} &
 \mc{1}{c|}{$D_5$} & \mc{1}{c|}{$A_4$} & \mc{1}{c|}{$A_2 \! \times \!
   A_1$} & \mc{1}{c|}{$\R \! \times \! A_1$} &
 \mc{1}{c|}{$\phantom{\Big(}{\R \atop A_1}\phantom{\Big)}$} &
 \mc{1}{c|}{$\{ e \}$} \\
\hline \hline
 D & \mc{1}{c|}{$n=8$} & \mc{1}{c|}{$n=7$} & \mc{1}{c|}{$n=6$} &
 \mc{1}{c|}{$n=5$} & \mc{1}{c|}{$n=4$} & \mc{1}{c|}{$n=3$} &
 \mc{1}{c|}{$n=2$} &\mc{1}{c|}{$n=1$} & \mc{1}{c|}{$n=0$} \\
\hline
\end{tabular}
\caption{The $E_8$ triangle} \label{mt}
\end{center}
\end{table}

This table of groups appearing in oxidation has a reflection symmetry
along the diagonal. This structure is easy to understand in th
present formalism\footnote{The proof of the symmetry given here is the
  one alluded to in a footnote in \cite{Henry-Labordere:2002dk}. It becomes
identical to the proof given in \cite{Henry-Labordere:2002dk}, upon
invoking one of the well known $A-D-E$ correspondences, between
Kleinian singularities and the classification of simply laced Lie
groups.}. Starting with a U-duality group $E_{n(n)}$ in 3-dimensions,
to find the U-duality group in $D$  dimensions, we decompose 
\be 
E_{n(n)} \rightarrow SL(D-2, \R) \times U_{n,D}
\ee
On the other hand, the $E$-series can be \emph{defined} as the groups
appearing in the following decomposition of regular subgroups 
\be
E_{8(8)} \rightarrow SL(9-n) \times E_{n(n)}
\ee
i.e. the $E_{n(n)}$ lattice is the orthogonal complement to the
$SL(9-n)$ lattice in $E_{8(8)}$. 

Combination of the 2 equations reveals that the groups
$U_{n,D}$ follow from the decomposition 
\be
E_8 \rightarrow SL(9-n, \R) \times SL(D-2,\R) \times U_{n,D}
\ee
There is an obvious symmetry, which is the symmetry of the magic
triangle:  
\be
U_{n,D}=U_{11-D,11-n}.
\ee

It is possible to define a triangle for every simply laced group $G$,
by decomposing 
\be 
G \rightarrow SL(n'-n, \R) \times SL(D-2,\R) \times U_{n,D},
\ee 
where $n'$ is defined by the largest $SL(n',\R)$ factor one can
find (it is the length of the largest chain in the extended Dynkin
diagram, plus 1). Such a triangle has the symmetry
\be
U_{n,D}=U_{n'+2-D,n'+2-n}
\ee
If we do not impose maximality on $n'$ the triangle will not be
symmetric, unless we impose a similar restriction on $D$. We have
renamed the original magic triangle to ``$E_8$''-triangle, to
distinguish it from other $A-D-E$-triangles.  

For non-simply laced groups the story is more complicated. The reason
is that we required the $SL(D \! - \! 2)$ of gravity to be made up from long
roots. To recover a symmetric triangle, we should demand the same for
the U-duality group. As illustration of the procedure, consider the
``triangle'' that we would obtain for $F_4$, depicted in table \ref{f4tab}. 

\begin{table}[ht]
\begin{center}
\begin{tabular}{|c|cccc}
\cline{1-2}
 6 & \mc{1}{|c|}{$\phantom{\Big(} A_1 \phantom{\Big)}$} & & & \\
\cline{1-2}
 5 & \mc{1}{|c|}{$\phantom{\Big(}A_2\phantom{\Big)}$} & & $\{e\}$ \\
\cline{1-3}
 4 & \mc{1}{|c|}{$\phantom{\Big(}C_3\phantom{\Big)}$} &
 \mc{1}{c|}{$C_2$} & $\R \times A_1$ & $\{ e \}$ \\
\cline{1-5}
 3 & \mc{1}{|c|}{$\phantom{\Big(}F_4\phantom{\Big)}$} &
 \mc{1}{c|}{$C_3$} & \mc{1}{c|}{$A_2$} & \mc{1}{c|}{$A_1$} \\ 
\hline
\hline
$D$ & \mc{1}{c|}{$n=4$} & \mc{1}{c|}{$n=3$} & \mc{1}{c|}{$n=2$} &
 \mc{1}{c|}{$n=1$} \\
\hline
\end{tabular}
\caption{The $F_4$ triangle}\label{f4tab}
\end{center}
\end{table}
The groups inside the boxes in the triangle are the ones one obtains
by restriction to level 1 subgroups. This looks odd, since $A_2$ and
$A_1$ appear in the $D=3$ row, and one surely would expect to be able
to decompactify from these. Hence we have added the groups that would
result from oxidation, but then evidently, the symmetry is gone. 

Summarizing, for the non-simply laced case the triangle either loses
its meaning as a table of groups encountered in oxidation, or it is
not symmetric. 

\subsection{The supergravity triangle}

There is a second ``magic triangle''. Again we put maximal
supergravity in the lower left corner, while going up represents
oxidation. To the right we list the theories obtained by truncating
the amount of 4-dimensional supersymmetries. The symmetries of the
theories obtained this way are displayed in the
tables \ref{mts} and \ref{mtsa}, which contain the information from a
table in \cite{Julia:1982gx} (an earlier version in
\cite{Julia:1980gr} has some small errors). Generically, the groups in
this triangle are non-split. We have listed along a vertical bar, two
characteristics of non-compact groups, being their signature (the
difference between the number of non-compact and the number of compact
generators, upper right corner), and their real rank (lower right
corner). In \cite{Arjannew, HJK} cosets of non-split groups will be
discussed in more detail. Table \ref{mtsa} lists the explicit forms
appearing in table \ref{mts}.  

This triangle has an intriguing approximate reflection symmetry in the
diagonal. The entries that exhibit the symmetry have groups that have
the same complexifications (but different real forms). With our theory
we can explain not only the approximate symmetry, but also why and
when it breaks down.

We need however a new ingredient, supersymmetry. It will be sufficient
to use some elements of the theory of spinors. Supersymmetric theories
enjoy (at the classical level) R-symmetries, mixing the various
supercharges. For supergravity theories, these R-symmetries are
(necessarily) local, and since they are compact, they must be embedded
inside maximal compact subgroups.  

The cosets reflect transverse degrees of freedom
\cite{Julia:1980gr}. Hence we are led to study spinors in $D-2$
Euclidean dimensions. Spinors are real, complex or quaternionic; they
are naturally acted upon by orthogonal, unitary, and symplectic
groups. We have listed the Clifford algebra's relevant for $D-2$
dimensions in table \ref{spin} (see e.g. \cite{Baez:2001dm} for a
recent review). We have also listed the symmetry
groups acting on $N$ supercharges, with the number $N$ referring to
the number of 4-dimensional supersymmetries. One of the things to note
is that the series of groups mentioned for $N=8$ are precisely the
compact subgroups for the $E_{n(n)}$ series; in this sense these groups
are ``tailor-made'' for maximal supergravity.  

\begin{table}[ht]
\begin{center}
\begin{tabular}{|c||c|c|c|}
\hline
 11 & $\R^{16}$ & $\{e\}$ & \\ \hline
 10 & $\R^8 \oplus \R^8$ & $O(1) \atop SO(2)$ & $\{ e \}$\\ \hline
 9 & $\R^8$ & $O(2)$ & $O(N/4)$ \\ \hline
 8 & $\C^4$ & $U(2)$ & $U(N/4)$ \\ \hline
 7 & $\h^2$ & $Sp(2)$ & $Sp(N/4)$ \\ \hline
 6 & $\h \oplus \h$ & $Sp(2) \! \times \! Sp(2)$ & $Sp(n) \! \times
 \!Sp(m)$\\ 
  & & & $ n+m = N/2, n,m \leq 2$ \\ \hline
 5 & $\h$ & $Sp(4)$ & $Sp(N/2)$ \\ \hline
 4 & $\C$ & $SU(8)$ & $U(N)$ \\ \hline
 3 & $\R$ & $SO(16)$ & $SO(2N)$ \\
\hline \hline
 D & Clifford & {$N=8$} & {$1 \leq N \leq 6$}\\
\hline
\end{tabular}
\caption{Clifford algebra's, and R-symmetries}\label{spin}
\end{center}
\end{table}

\begin{table}
\small
\begin{center}
\begin{tabular}{|c|cccccccc}
\cline{1-1}
 11 & \mc{1}{|c}{$\{ e \} \big|^0_0$} & \phantom{\Big|} & & & & & & \\
\cline{1-2}
 10 & \mc{1}{|c|}{$\R, A_{1}\big|^1_1$} & \phantom{\Big|} & & & & & &
 \\ 
\cline{1-2}
 9 & \mc{1}{|c|}{$\R \! \times \! A_{1} \big|^2_2$} & \phantom{\Big|}
 & & & & & & \\ 
\cline{1-2}
 8 & \mc{1}{|c|}{$A_{1} \! \times \!A_{2} \big|^3_3$} &
 \phantom{\Big|} & & & & & & \\ 
\cline{1-2}
 7 & \mc{1}{|c|}{$A_{4}\big|^4_4$} & \phantom{\Big|} & & & & & & \\
\cline{1-3}
 6 & \mc{1}{|c|}{$D_{5}\big|^5_5$} & \mc{1}{c|}{$A_{1} \! \times \!
 A_{3}\big|^{\textrm{-}8}_1$} & \phantom{\Big|} & & & & & \\ 
\cline{1-3}
 5 & \mc{1}{|c|}{$E_{6}\big|^6_6$} &
 \mc{1}{c|}{$A_{5}\big|^{\textrm{-}7}_2$} & & \phantom{\Big|} & & & \\
\cline{1-5}
 4 & \mc{1}{|c|}{$E_{7}\big|^7_7$} &
 \mc{1}{c|}{$D_{6}\big|^{\textrm{-}6}_3$} &
 \mc{1}{c|}{$A_{5}\big|^{\textrm{-}15}_1$} & \mc{1}{c|}{$A_{1} \!
 \times \! A_{3}\big|^{\textrm{-}14}_1$} & $U(3)
 \big|^{\textrm{-}9}_0$ & $U(2)\big|^{\textrm{-}4}_0$ &
 $U(1)\big|^{\textrm{-}1}_0$ & $\phantom{\Big|}\! \{e\}\big|^0_0$ \\
\cline{1-9}
 3 & \mc{1}{|c|}{$E_{8}\big|^8_8$} &
 \mc{1}{c|}{$E_{7}\big|^{\textrm{-}5}_4$} &
 \mc{1}{c|}{$E_{6}\big|^{\textrm{-}14}_2$} &
 \mc{1}{c|}{$D_{5}\big|^{\textrm{-}13}_2$} &
 \mc{1}{c|}{$A_{4}\big|^{\textrm{-}8}_1$} & \mc{1}{c|}{$A_{2} \!
 \times \! A_{1}\big|^{\textrm{-}3}_1$} & \mc{1}{c|}{$U(1) \! \times
 \! A_{1}\big|^0_1$} & \mc{1}{c|}{$\phantom{\Big|} \! A_{1}\big|^1_1$}
 \\
\hline \hline
 D & \mc{1}{c|}{$N=8$} & \mc{1}{c|}{$N=6$} & \mc{1}{c|}{$N=5$} &
 \mc{1}{c|}{$N=4$} & \mc{1}{c|}{$N=3$} & \mc{1}{c|}{$N=2$} &
 \mc{1}{c|}{$N=1$} &\mc{1}{c|}{$N=0$} \\
\hline
\end{tabular}
\caption{The supergravity triangle} \label{mts}

\vspace{0.5cm}

\normalsize

\begin{tabular}{|cc|c|c|c|}
\hline
 $N$ & $D$ & group & real form & compact \\
\hline
 6 & 6 & $A_{1} \! \times \! A_{3}\big|^{\textrm{-}8}_1$ & $Sp(1)
 \times SU^*(4)$ & $Sp(1) \times Sp(2)$ \\
 6 & 5 & $A_{5}\big|^{\textrm{-}7}_2$ & $SU^*(6)$   & $Sp(3)$ \\
 6 & 4 & $D_{6}\big|^{\textrm{-}6}_3$ & $SO^*(12)$  & $U(6)$  \\
 6 & 3 & $E_{7}\big|^{\textrm{-}5}_4$ & $E_{7(-5)}$ & $SU(2) \times
 SO(12)$\\ 
\hline 
 5 & 4 & $A_{5}\big|^{\textrm{-}15}_1$ & $SU(5,1)$ & $U(5)$  \\
 5 & 3 & $E_{6}\big|^{\textrm{-}14}_2$ & $E_{6(-14)}$ & $U(1) \times
 SO(10)$ \\ 
\hline
 4 & 4 & $A_{1} \! \times \! A_{3}\big|^{\textrm{-}14}_1$ & $SL(2)
 \times SU(4)$ & $U(4)$ \\
 4 & 3 & $D_{5}\big|^{\textrm{-}13}_2$ & $SO(8,2)$ & $U(1) \times
 SO(8)$ \\ 
\hline
 3 & 3 & $A_{4}\big|^{\textrm{-}8}_1$  & $SU(4,1)$ & $U(1) \times
 SO(6)$ \\ 
\hline
 2 & 3 & $A_{2} \! \times \! A_{1}\big|^{\textrm{-}3}_1$ & $SU(2,1)
 \times SU(2)$ & $U(1) \times SO(4)$ \\
\hline
 1 & 3 & $U(1) \! \times \! A_{1}\big|^0_1$ & $U(1) \times SL(2,\R)$ &
 $U(1) \times SO(2)$ \\
\hline
 0 & 3 & $A_{1}\big|^1_1$ & $SL(2,\R)$ & $SO(2)$ \\
\hline
\end{tabular}
\caption{Addendum to the supergravity triangle} \label{mtsa}
\end{center}
\end{table}

The supergravity triangle \ref{mts} is based on the number of
supersymmetries in 4 dimensions. In 4 dimensions, the helicity group
is $O(2)$. There are two transverse dimensions, and spinors in two
dimensions are complex. Hence the R-symmetry group for
N supersymmetries must contain $U(N)$ with $N$ the number of
supersymmetries \cite{Cremmer:1977zt}. Exception to the rule is
$N=7$. As is well known, building a supergravity multiplet with $7$
supersymmetries, one automatically finds an eighth supersymmetry,
because a realistic theory has to contain the CPT-conjugate, and the
$N=7$ multiplet and its conjugate automatically fill a single $N=8$
multiplet. Then for $N=8$, the supergravity multiplet is its own conjugate
(in 4-d). The $U(1)$ factor in $U(N)$ multiplies a multiplet with a
phase, and the CPT-conjugate with the opposite phase. If a multiplet is
its own CPT-conjugate, then $U(1)$ must act trivially on the multiplet,
and since there is no other multiplet for $N=8$ than the supergravity
multiplet, the R-symmetry is truncated to $SU(8)$ (see also
\cite{Fre:2001jd}).   

Now suppose we truncate to fewer supersymmetries, say $n$. We
decompose the $N=8$ multiplet, by separating the supercharges in a
group of $n$, and one of $8-n$ supercharges. The $8-n$ supercharges
have $SU(8\!-\!n)$ symmetry, and the remaining ones $U(n)$. We should
therefore decompose via 
\be 
SU(8) \rightarrow U(n) \times SU(8\!-\!n)
\ee 

In 4-d maximal supergravity, the U-duality group is $E_{7(7)}$, and
its maximal compact subgroup is $SU(8)$. Truncating to fewer
supersymmetries means truncating the R-symmetry group by an $SU(8\!-\!n)$
factor, and because $SU(8)$ is the maximal level 1 compact subgroup,
it means truncating $E_{7(7)}$ by a regular $SU(8\!-\!n)$ factor.  

From the previous section we already knew that $E_{7(7)}$ itself is
obtained by decomposing $E_{8(8)}$ with an $SL(2)$ factor. We can now
move in two directions: Fewer (four-dimensional) supersymmetries; and,
more dimensions. For the U-duality group $U_{N,D}$ of a theory in $D$
dimensions, with $N$ supersymmetries as counted in 4 dimensions, we
decompose 
\be \label{mtsugra}
E_{8(8)} \rightarrow SU(8\!-\!N) \times SL(D\!-\!2,\R) \times U_{N,D},
\ee
where $SU(8\!-\!N)$ and $SL(D\!-\!2)$ are \emph{regular} (level 1)
subgroups. Note that for $N=7$ we get the groups of $N=8$
supersymmetry; the formula reflects that we get an extra
supersymmetry for free. 

To exhibit the symmetry, we replace the groups of (\ref{mtsugra}) by
their complexifications:
\be \label{cmplxmtsugra}
E_8 \rightarrow A_{7-N} \times A_{D-3} \times \widetilde{U}_{N,D}
\ee
It is clear that $\widetilde{U}_{N,D} =
\widetilde{U}_{10-D,10-N}$. This is responsible for the approximate
symmetry of the supergravity triangle. The symmetry would be exact, if the
symmetry extended from $\widetilde{U}_{N,D}$ to $U_{N,D}$, but it is
easy to see that it does not. For example, the maximal $SL(D \! - \! 2)$
subgroup in $E_{8(8)}$ is $SL(9, \R)$ giving 11-d supergravity, while
the maximal $SU(8\!-\!N)$ group is $SU(8)$ (sitting inside $SO(16)$),
resulting in no supersymmetry.  

The last example leads to the decomposition
\be
\frac{E_{8(8)}}{SO(16)} \rightarrow \frac{E_{7(7)}}{SU(8)} \times
\frac{SL(2)}{SO(2)} 
\ee 

It is amusing to see that in this way, the maximal supersymmetric
theory delivers us the correct Ehlers symmetry of 4-d
non-supersymmetric gravity, compactified to 3-d. It appears next to
the U-duality group of maximal supergravity in 4-d.

\section{Discussion and conclusions}\label{concl}

The central theme of this paper is that the inverse process to
dimensional reduction, oxidation, is completely governed by
group theory. We have restricted to oxidation starting from coset
theories in 3 dimensions, where the groups involved are split. In
follow-up papers \cite{Arjannew, HJK} the formalism will be extended to all
non-compact forms, and results will be presented for the oxidation chains for
these groups, in particular their maximal dimension. The analysis will
require a few extra technicalities, but still be completely
based on group theory.

The final oxidation recipe is very simple: It can be found in section
\ref{inter}, and described in a few lines. Note that it defines the
oxidized theory in terms of equations of motion and Bianchi
identities; in particular, there is no difficulty describing theories
with self-dual forms. Our formalism is democratic in its equations; it
does not prescribe which one is supposed to be a dynamical equation, and
which one a constraint equation. As such, we are also insensitive to
the modifications of the symmetries resulting from dualizing various
forms \cite{Lu:1997df, Cremmer:1997ct}; these affect the actions, and
therefore the interpretation of the equations, but not the equations
themselves. 

We started with 3 dimensional theories. This has a number of profound
consequences: The groups involved are finite dimensional, and the relevant 
representation theory is relatively simple. We believe that there are
no fundamental obstacles in relating the theories in this paper to 2-
and lower dimensional theories, but evidently, this should involve the
representation theory of infinite dimensional groups, and therefore be
more involved.

There are a number of previous developments that deserve some
comments. 

A paper that develops a number of directions similar to the ones
expressed in this paper, yet differs profoundly in its philosophy, is
\cite{Cremmer:1998px}. In this paper, the fields of maximal
supergravities are supplemented with their ``double'' (essentially the
Poincar\'e dual). This leads to doubled Lagrangians, with twice the
amount of fields, from which the original formulation can be rederived
by imposing a ``twisted self-duality'' equation. For axions and form
fields, the ``doubles'' of \cite{Cremmer:1998px} obey Bianchi
identities that are equivalent to our equations of motion. An important
difference between our work and \cite{Cremmer:1998px}, is that in the
latter paper also the dilatons are doubled, a procedure that has no
analogue in our paper. This seems to be the source of another key
difference: \cite{Cremmer:1998px} finds a particular kind of
supergroups, while we have claimed that ``ordinary'' Lie groups can
account for all the structure. Note also an important difference in
the treatment of Chern-Simons terms: In \cite{Cremmer:1998px} these
are an ingredient in deforming the superalgebra, whereas in our work,
we have ignored them for a long while, finding in the end that they
come out automatically. In view of the flexibility and content of both
formalisms, it seems that there should be some well-defined map between
them. The supergroups formed one of the ingredients for the
development in \cite{Henry-Labordere:2002dk}. 

Another intriguing paper \cite{Iqbal:2001ye} suggests relations
between the classification of Del Pezzo surfaces, and maximal
supergravities in various dimensions. The work
\cite{Henry-Labordere:2002dk} elaborates on the theme, and relates it
to the supergroups of \cite{Cremmer:1998px}. Many of the relations
found are linked to the fact that the second homology lattice
of the Del Pezzo's are the root lattices of the $E_n$ groups. Our work
emphasizes the physical information that can be extracted from the
root lattices. It seems fair to say that the Del Pezzo-supergravity
correspondence requires more motivation from the physical side, as
there are clear links between supergravity and $E_n$ groups, as well
as between Del Pezzo's and $E_n$ groups, but it is not clear to what
extent there is a third link of a potential triangle, between Del
Pezzo's and supergravities. Much of the information uncovered
in \cite{Iqbal:2001ye,Henry-Labordere:2002dk} might actually be
recoverable from group theory (certainly the BPS-states should have
some place; we have all the bosonic equations of the theory), and a
direct physical motivation for the involvement of Del Pezzo's seems to
be still lacking (apart from some remarks in \cite{Iqbal:2001ye}). On the
other hand, \emph{if} there is a deeper reason for a Del Pezzo-gravity
correspondence, the present formalism might provide valuable tools for
exploring it. 

An old idea is that the symmetries that (maximal) supergravities
exhibit upon compactification, are in some way already realized in the
higher dimensional theories \cite{Julia:1980gr}. This conjecture in
its most advanced form is at the heart of the $E_{11}$-program of
\cite{West:2001as}. Our paper fits into the strategy of identifying
substructures of $E_{11}$: We have demonstrated how to reconstruct the
bosonic sector of 11 dimensional supergravity (and its
compactifications) from the finite dimensional algebra $E_8$. We
believe that its very plausible that an extension of our methods to
compactifications lower than 3 dimensions exists (the recent paper
\cite{Damour:2002cu} has partial results on an $E_{10}$ coset which
seem reminiscent to ours for finite dimensional cosets) and that such
an extension might lead to support, and perhaps even a proof of the
conjecture of \cite{West:2001as}. 

Some papers are hinting at evidence for a ``$12^{th}$ dimension'' for
maximal supergravities (see \cite{Obers:1998fb, Cremmer:1998px,
  Vafa:1996xn} for some motivations). We have nothing to say in
support of this conjecture, but do note that restrictions and
assumptions that formed the basis of this paper made it unlikely that
we would be able to probe such a dimension. The restrictions imposed
by supersymmetry require that a $12^{th}$ dimension be very different
in nature from the other 11 space-time dimensions, which in turn seems
incompatible with our methods, which are (indirectly) rooted in
Lorentz symmetries. 

Similarly, although our methods are easily applicable to the massless
sector of the bosonic string, (governed by the split form of $D_{24}$,
$SO(24,24)$), we find no hints for ``bosonic M-theory''
\cite{Horowitz:2000gn}. Again it can be argued that our assumptions
are not suitable for dealing with this issue, but it is less clear to
us why (from a physical perspective) this hypothetical theory should
not fit somewhere in our framework (it cannot fit anywhere by virtue
of the classification of simple Lie algebra's).

After these negative results we would like to stress once more a
remarkable positive result. We have found novel branches for the $B_n$
and $D_n$ series in 6 dimensions. This is remarkable, because these
theories oxidize in their maximal dimension to a theory containing
gravity, an antisymmetric 2-tensor, and a dilaton scalar (the $B_n$
series, and the non-split forms \cite{Arjannew, HJK} have extra vectors);
at the classical level, they allow elementary string solutions, and
hence are string theories. In 6 dimensions, we find
gravity, scalars, and multiple antisymmetric tensors (for the $B_n$
series and the non-split forms \cite{Arjannew, HJK}, the number of
self-dual tensors is not equal to the number of antisymmetric tensors,
and the theories are chiral, even when fermions are absent). The only
known theory in this category is a branch of the $D_8$ chain, giving
an extension of the bosonic sector of $(2,0)$ gravity with tensor
multiplets (which can be obtained by compactifying $IIB$ gravity on a
$K3$, and truncating; $IIB$ on $K3$ itself gives rise to a non-split
form of $D_{16}$). The other ones, in particular the
theory arising as a branch on the $D_{24}$ oxidation chain of the
massless sector of the bosonic string, represent new challenges to
those who believe that algebraic structure must lead to meaningful
physics.

Apart from the above problems there are many other perspectives for
future work. We will extend our analysis to cosets of non-split
groups \cite{Arjannew, HJK}. Some fundamental directions for future
research are the relation with supersymmetry, and the
extension to theories that have less than 3 non-compact dimensions. We
also believe that there is more to say on the microscopic realization
of the symmetries. On the other hand, we hope that this work may also
lead to applications. Possible directions are the investigation of
gaugings of (super)gravities, and the study of solutions of the
theories. 

{\bf Acknowledgements:} I would especially like to thank Bernard Julia
for discussions, encouragement, and a critical reading of the draft. A
project with A.~Hanany and B.~Julia \cite{HJK} inspired me to research
the issues presented here. I also thank Niels Obers, Pierre
Henry-Labordere and Louis Paulot for discussions. This work was
started at the E.N.S., Paris, where the author was supported by EU
contract HPRN-CT-2000-00122. The author is supported in part by the
``FWO-Vlaanderen'' through project G.0034.02, in part by the Federal
office for Scientific, Technical and Cultural Affairs trough the
Interuniversity Attraction Pole P5/27 and in part by the European
Commision RTN programme HPRN-CT-2000-00131, in which the author is
associated to the University of Leuven. 

\appendix

\section{Conventions for Lie algebra's}\label{Lie}

We deal exclusively with real Lie algebra's. We follow to
some extent (but not completely), the conventions of
\cite{Cornwell}. This reference also discusses some
alternative conventions.

Every real Lie algebra $\cal L$ has an adjoint representation. With a
basis $b_i$ for the Lie algebra, the matrices of the adjoint
representation are defined by 
\be  
\{ \textrm{ad}(a)\}_{ij}\ b_j= [a,b_i] \qquad a,b, \in L
\ee

The Killing form, which we denote by $\inp{\ . \ }{ \ .\ }$, is
defined by
\be
\inp{a}{b}= \textrm{tr}(\textrm{ad}(a)\textrm{ad}(b)) \qquad a,b \in L
\ee 
This fixes many properties of the Killing form, but not its
normalization. In our paper the Killing form defines the inner product
on the scalar manifold spanned by the dilatons, and hence its
normalization corresponds to the normalization of the dilatons.

The Cartan subalgebra $\cal H$ of $\cal L$ is a maximal Abelian
subalgebra, with as additional property that $\textrm{ad}\cal H$ is
completely reducible. The dimension $r$ of the Cartan subalgebra is
called the rank of the algebra. We choose an orthonormal basis
$H_1,\ldots, H_r$ of $\cal H$:
\be \label{kilcar}
\inp{H_i}{H_j}=\delta_{ij}
\ee

As the $\textrm{ad}(H_i)$ are mutually commuting, they can
be simultaneously diagonalized. The simultaneous eigenvectors for the 
$\textrm{ad}(H_i)$, corresponding to non-zero eigenvalues are denoted
by $E_{\alpha}$. The commutation
relations following from these definitions are
\be
\left[ H_i, H_j \right] = 0 \qquad \left[ H_i, E_{\alpha} \right] =
\alpha_i E_{\alpha} 
\ee
It is common to regard the $\alpha_i$ as components of a
vector. Contracting them with the basis elements of $\cal H$, we define
$\alpha = \alpha_i H_i$. We have $\inp{\alpha}{\beta}= \alpha_i
\beta_i$, like a normal inner product. The $\alpha$'s appearing as
label on $E_{\alpha}$ are called roots. Inspection of the simple Lie
algebra's shows that the norm of the roots can take at most two
values, hence one speaks of short and long roots. We normalize, as
conventional, the long roots to have length
$\sqrt{2}$, but note that this leads to an unconventional dilaton kinetic
term in the matter actions in this paper. The normalization of
$\inp{E_{\alpha}}{E_{-\alpha}}$ is not yet determined, and we set
\be \label{killad}
\inp{E_{\alpha}}{E_{-\alpha}}=1
\ee
That this norm is independent of $\alpha$ is important for the
normalization of the axion kinetic terms in the paper. Combinations of
two generators inserted into the Killing form that cannot be brought
to the form (\ref{kilcar}) or (\ref{killad}) are zero.
 
The remaining commutation relations are
\be 
\left[ E_{\alpha}, E_{-\alpha} \right] = \alpha_i H_i \qquad 
\left[ E_{\alpha}, E_{\beta} \right] = N_{\alpha,\beta} E_{\alpha+
  \beta}, \ \alpha + \beta \neq 0
\ee
In the last of these commutation relations, the structure constants
$N_{\alpha,\beta}$ appear, that are non-zero whenever $\alpha +
\beta$ is a root. They are antisymmetric $N_{\beta,\alpha}= -
N_{\alpha,\beta}$, and may be chosen such that
$N_{\alpha,\beta} = -N_{-\alpha,-\beta}$. Further properties are \cite{Cornwell}
\bea
N_{\alpha,\beta} = N_{\beta,\gamma} = N_{\gamma, \alpha} & & \textrm{ if
}\alpha + \beta + \gamma =0 \label{strcycl} \\
N_{\alpha,\beta} N_{\gamma, \delta} +N_{\beta, \gamma} N_{\alpha,
  \delta} +N_{\gamma, \alpha} N_{\beta, \delta}= 0 & & \textrm{ if
}\alpha + \beta + \gamma + \delta =0 \label{strjacobi}
\eea    
For completeness (though we hardly will use it) we mention that
\be 
\{ N_{\alpha,\beta} \}^2 = \hlf \inp{\alpha}{ \alpha} q(p+1)
\ee
with $p$ and $q$ poitive integers such that the $\alpha$-string
containing $\beta$ 
is $\beta-p\alpha, \ldots\beta, \ldots, \beta+ q\alpha$
\cite{Cornwell}. These conventions do not fix all signs, the remaining
ones may be chosen freely.

The groups in this paper are in split form. We define the split form
as the form that is generated by combinations of the $H_j$ and
$E_{\alpha}$ with real coefficients (while for example the compact
form has generators like $iH_j$, $i(E_{\alpha}+E_{-\alpha}$)). The
maximal compact subgroup of the split form
is generated by generators of the form $E_{\alpha}-E_{-\alpha}$. The
compact subgroup can be 
used to generate Weyl reflections on the root space, so, as in the
case of compact groups, different labellings of the roots related by
Weyl reflections are equivalent. 

The set of roots is denoted by $\Delta$. One can choose a basis
for the root lattice consisting of simple roots. The simple
roots $\alpha_i$ are characterized by the property, that
if $\alpha_i$ and $\alpha_j$ are simple roots, then $\alpha_i
-\alpha_j \notin 
\Delta$. The roots can be expanded in the simple roots, as 
$$
\alpha = \sum_i p_i \alpha_i
$$
The coefficients $p_i$ are integers, that are either all non-negative
or all non-positive; consequently the non-zero roots can be 
divided in positive and negative roots. The set of positive roots is
denoted by $\Delta^+$. We also define the height $h$ of the root
$\alpha$ by, $h = \sum_i p_i$; for a non-zero root this a non-zero number. 

The fundamental weights are defined by
\be
2 \frac{\inp{\alpha_i}{\lambda_j}}{\inp{\alpha_i}{\alpha_i}} =\delta_{ij}
\ee
The weight lattice is spanned by the fundamental weights. Also to the
weights one can attach a height, by expanding them in the simple
roots, and summing up the coefficients in the expansion (that need not
be integer). A standard result is that each irreducible representation
is characterized by a unique highest weight.

The algebra defined by the $H_i$ and $E_{\alpha}$ allows the Chevalley
involution $\omega$, with action 
\be
\omega: H_i \rightarrow -H_i \qquad E_{\alpha} \rightarrow -
E_{-\alpha} 
\ee
The significance here of this involution lies in the
fact that the generators invariant under the involution are of the
form $E_{\alpha}-E_{-\alpha}$, and generate the compact
subgroup. Specifically, for the algebra of $SL(n,\R)$, the invariant
generators are the generators of $SO(n)$. Given an algebra element
$T$, we define $T^{\#}= -\omega(T)$. For the fundamental representation
of $SL(n,\R)$ the generators can be chosen such that $T^{\#}$ acts as
matrix transpose. The generators invariant under $\omega$ satisfy
$T^{\#}= -T$, and hence are antisymmetric. This is the root of the name
``generalized transpose'' used in \cite{Cremmer:1997ct} and the
present paper. It is however possible to choose the matrix
representation of the adjoint representation such that the
``generalized transpose'' \emph{is} actually the transpose (be it in a
much bigger representation). It is therefore possible to extend the
action 
of the generalized transpose to the group, by setting $(\exp A)^{\#} =
\exp (A^{\#})$, a property obviously satisfied by the ``normal''
transpose. The invariant subgroup is then defined by $O^{\#} O =1$.

\section{Oxidation for cosets of split forms of Lie groups} \label{coset}

\begin{figure}[ht]
\begin{center}
\includegraphics[width=10cm]{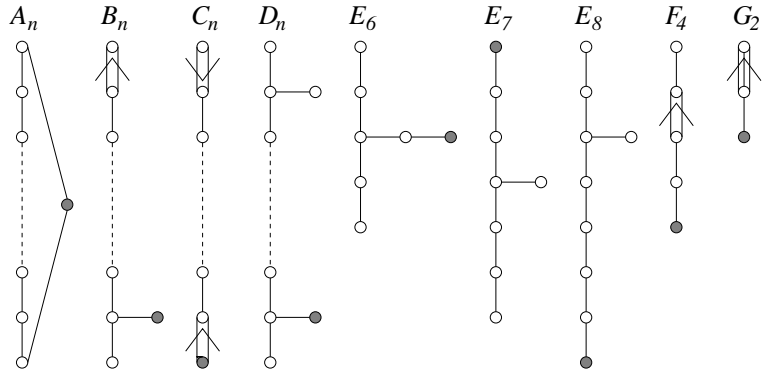}
\caption{Extended Dynkin diagrams: The marked dot is the extended
  node. Erasing this node and all its connections leaves the standard
  Dynkin diagram.}  \label{dynkfig}
\end{center}
\end{figure}

In this appendix we list the decompositions of split Lie
groups into the \mbox{$SL(D \! - \! 2) \times U_D$} groups that are
central to this paper. The table has been compiled with the help of
\cite{McKay}. We also list all extended Dynkin diagrams, in figure
\ref{dynkfig}

We use the notations $\{ e\}$ and $SL(1,\R)$ for the trivial group,
signifying triviality of the U-duality group, or that we are dealing
with the 3-dimensional theory, respectively. A number in boldface
denotes the dimension of the representation; the conjugate
representation is denoted by putting a bar over the number. The
representations of the Abelian groups ($\R$ factors) are labeled by their
charges. They are normalized such that charge 1 would correspond to
the minimal charge in the compact, $U(1)$-case. The listed charges
therefore reflect the ratios rather than absolute charges. At the
level of the equations of motion and Bianchi identities, the unit of
charge can be absorbed in a redefinition of the corresponding
dilaton. In the action however this may lead to an unconventional
kinetic term for the dilaton. 

With the contents of this appendix and the formalism developed in the
text, the reader may rederive the results of \cite{Cremmer:1999du},
and fill in some small omissions in that paper (already mostly filled
in in \cite{Henry-Labordere:2002dk}). The list of representations in
this appendix is complete in the sense that all relevant level 1
embeddings are given here, and hence, under the assumptions that lead
us to restrict to level 1 subgroups, this
appendix lists all the possible theories that can be oxidized from a
given split simple Lie group. In \cite{Arjannew} we will complete this
project by setting up the theory for the non-split Lie groups. 

The paper \cite{Cremmer:1999du} discusses in detail the theories
appearing here, in 3 dimensions and in their maximal dimension, and
sometimes in 4 dimensions. Other references are given in the relevant
subsections. 

\subsection{$A_n$}

The split real form of $A_n$ is $SL(n+1, \R)$. These theories oxidize to
pure gravity in $n+3$ dimensions. The reduction of these theories is
discussed in detail in \cite{Cremmer:1998px}. 

The table of decompositions is ($d+d'=n$):
\bd
\ba{l@{}cl}
SL(n,\R) \! \times \! \{e \}    & : & \mathbf{(n^2\!-\!1)} \\
  \vdots     &   & \vdots \\
SL(d,\R) \! \times \! SL(d', \R) \! \times \! \R &: &
\mathbf{(d^2\!-\!1,1)^0 \! \oplus \! 
  (1,d'^2\!-\!1)^0 \! \oplus \! (1,1)^0  \oplus }\\ 
 & &  \mathbf{ (d,d')^n \! \oplus \! (\ov{d}, \ov{d'})^{-n}} \\
  \vdots     &   & \vdots \\
 SL(1,\R) \! \times \! SL(n,\R)    & : & \mathbf{(n^2\!-\!1)} 
\ea
\ed

\subsection{$B_n$}

The split real form of $B_n$ is $SO(n+1,n)$. Generically, these
oxidize to $n+2$ dimensions, and then contain gravity, an antisymmetric
2-tensor, a dilaton, and a vector. This matter content is typical for
string theories; these theories are indeed included in the $SO(d+r,d)$
theories following from general considerations on toroidally
compactified strings \cite{Narain:1985jj}. The $B_8$ case allows a
supersymmetric extension. The $B_2$ case is interesting, because it is
the minimal simple group that leads to (classical) S-duality in 4
dimensions. The $B_3$ case oxidizes to gravity in 6 dimensions coupled
to a single self dual tensor. All $B_n$-theories with $n>3$ have a
separate branch in 6 dimensions, and all theories with $n \geq 2$ have
a separate $SL(2,\R)$ factor in 4 dimensions, signifying S-duality.

The table of decompositions is ($d+d'=n$):
\bd
\ba{l@{}cl}
SL(n, \R) \! \times \! \R  & : &\mathbf{(n^2\!-\!1)^0 \! \oplus \! 1^0
  \! \oplus \! n^1 \! \oplus \!  \frac{n(n-1)}{2}^2 \! \oplus \!
  \ov{\frac{n(n-1)}{2}}^{-2} \! \oplus \! \ov{n}^{-1}} \\ 
 \vdots & & \vdots \\ 
SL(d,\R) \! \times \! SO(d',d'+1) \! \times \! \R & : &
\mathbf{(d^2\!-\!1,1)^0  \! \oplus \! (1,d'(2d'+1))^0 \! \oplus \!
  (1,1)^0 \oplus }\\  
 & & \mathbf{(d,2d'+1)^1 \! \oplus \! (\ov{d},2d'+1)^{-1} \oplus }\\
 & & \mathbf{(\frac{d(d-1)}{2},1)^2 \! \oplus \!
  (\ov{\frac{d(d-1)}{2}},1)^{-2}} \\  
 \vdots & & \vdots \\ 
SL(4,\R) \! \times \! SO(n\!-\!3,n\!-\!4) \! \times \! \R & : &
\mathbf{(15,1)^0 \! \oplus \! (1,(n\!-\!4)(2n\!-\!7))^0 \! \oplus \!
  (1,1)^0 \oplus }\\  
 & & \mathbf{(4,2n\!-\!7)^1 \! \oplus \! (\ov{4},2n\!-\!7)^{-1} \!
  \oplus \! (6,1)^{2} \! \oplus \! (6,1)^{-2}} \\
SL(3,\R) \! \times \! SO(n\!-\!2,n\!-\!3) \! \times \! \R & : &
\mathbf{(8,1)^0 
  \! \oplus \! (1,(n\!-\!3)(2n\!-\!5))^0 \! \oplus \! (1,1)^0 \oplus }\\ 
 & & \mathbf{(3,2n\!-\!5)^1 \! \oplus \! (\ov{3},2n\!-\!5)^{-1} \!
  \oplus \! (3,1)^{-2} \! \oplus \!  (\bar{3},1)^2} \\ 
SL(2,\R) \! \times \! SO(n\!-\!1,n\!-\!2) \! \times \! SL(2,\R) & : &
\mathbf{(3,1,1) \! \oplus \! (1,(n\!-\!2)(2n\!-\!3),1) \oplus }\\ 
 & & \mathbf{(1,1,3) \! \oplus \! (2,2n\!-\!3,2)} \\
SL(1,\R) \! \times \! SO(n\!+\!1,n) & : & \mathbf{n(2n+1)}
\ea
\ed

There exists an alternative decomposition with a maximal $SL(4,\R)$
factor   
\bd
\ba{l@{}cl}
SL(4,\R) \! \times \! SO(n\!-\!2,n\!-\!3) & : & \mathbf{(15,1) \!
  \oplus \! (1,(n\!-\!3)(2n\!-\!5)) \! \oplus \! (6,2n\!-\!5)}  
\ea
\ed

The table also gives the correct results for $4 \geq n > 1$, provided
one restricts to the lines where the number $p$ in the always present
$SO(p+1,p)$ factor is bigger or equal than $0$ ($SO(1,0)$ is the
trivial group), and omits representations that would be $0$-dimensional
according to the above formulas. 

We have $SO(2,1) \cong SL(2,\R)$
and this theory can be found in the previous paragraph.  

For $n=3$ the ``extra'' line is possible, while the one in the table
for $SL(4,\R)$ is not.

\subsection{$C_n$}

The split real form of $C_n$ is $Sp(n,\R)$. These theories all oxidize
to 4 dimensions, no matter what the value of $n$ is. The $C_2 \cong
B_2$ theory has S-duality in 4 dimensions. The $C_n$ theories with $n
> 2$ have an extended version of S-duality, involving a higher
dimensional $Sp(n-1,\R)$ group. 

The table of decompositions is:
\bd
\ba{l@{}cl}
SL(2,\R) \! \times \! Sp(n\!-\!1,\R) & : & \mathbf{(3,1) \! \oplus \!
  (1,(n\!-\!1)(2n\!-\!1)) \! \oplus \! (2,2n\!-\!2)} \\ 
SL(1,\R) \! \times \! Sp(n,\R) & : & \mathbf{n(2n+1)}
\ea
\ed
Note that for $n=2$ we have $Sp(2, \R) \cong SO(3,2)$ which confirms
the previous analysis, while for $n=1$ we have $Sp(1,\R) \cong
SL(2,\R)$. 

\subsection{$D_n$}

The split real form of $D_n$ is $SO(n,n)$. These theories generically
oxidize to $n+2$ dimensions, and contain gravity, an antisymmetric
2-tensor, a dilaton, again typical for string theories. Indeed, they
are included in the $SO(d+r,d)$ theories following from general
considerations \cite{Narain:1985jj}. 
 The $D_8$ case allows a supersymmetric extension, and
corresponds then to pure type I supergravity in 10 dimensions
\cite{Chamseddine:ez}. The $D_3$ case oxidizes to pure gravity in 6
dimensions. All $D_n$-theories with $n>3$ have a separate branch in 6
dimensions, and all theories with $n \geq 2$ have a separate
$SL(2,\R)$ factor in 4 dimensions, signifying S-duality. The $D_4$
theory can be oxidized to 6 dimensions; compactifying this theory on a
circle it displays a triality rather than a duality.

The table of decompositions is:
\bd
\ba{l@{}cl}
SL(n, \R) \! \times \! \R  & : &\mathbf{(n^2\!-\!1)^0 \! \oplus \! 1^0
  \! \oplus \! 
  \frac{n(n-1)}{2}^2 \! \oplus \! \ov{\frac{n(n-1)}{2}}^{-2}} \\ 
 \vdots & & \vdots \\
SL(d,\R) \! \times \! SO(d',d') \! \times \! \R & : &
\mathbf{(d^2\!-\!1,1)^0 \! \oplus \! (1,d'(2d'\!-\!1))^0 \! \oplus \!
  (1,1)^0 \oplus }\\  
 & & \mathbf{(d,2d')^1 \! \oplus \! (\ov{d},2d')^{-1} \oplus }\\
 & & \mathbf{(\frac{d(d-1)}{2},1)^2 \! \oplus \!
  (\ov{\frac{d(d-1)}{2}},1)^{-2}} \\ 
 \vdots & & \vdots \\
SL(4,\R) \! \times \! SO(n\!-\!4,n\!-\!4) \! \times \! \R & : &
\mathbf{(15,1)^0 \! \oplus \! (1,(n\!-\!4)(2n\!-\!9))^0 \! \oplus \!
  (1,1)^0 \oplus }\\ 
 & & \mathbf{(4,2n\!-\!8)^1 \! \oplus \! (\ov{4},2n\!-\!8)^{-1} \!
  \oplus \! (6,1)^2 \! \oplus \! (6,1)^{-2}} \\  
SL(3,\R) \! \times \! SO(n\!-\!3,n\!-\!3) \! \times \! \R & : &
\mathbf{(8,1)^0 
  \! \oplus \! (1,(n\!-\!3)(2n\!-\!7))^0 \! \oplus \! (1,1)^0 \oplus }\\ 
 & & \mathbf{(3,2n\!-\!6)^1 \! \oplus \! (\ov{3},2n\!-\!6)^{-1} \!
  \oplus \! (3,1)^{2} \! \oplus \! 
  (\bar{3},1)^{-2}} \\ 
SL(2,\R) \! \times \! SO(n\!-\!2,n\!-\!2) \! \times \! SL(2,\R) & : &
\mathbf{(3,1,1) \! \oplus \! (1,(n\!-\!2)(2n\!-\!5),1) \oplus }\\ 
 & & \mathbf{(1,1,3) \! \oplus \! (2,2n\!-\!4,2)} \\
SL(1,\R) \! \times \! SO(n,n) & : & \mathbf{n(2n\!-\!1)}
\ea
\ed

There exists an alternative decomposition with a maximal $SL(4,\R)$
factor  
\bd
\ba{l@{}cl}
SL(4,\R) \! \times \! SO(n\!-\!3,n\!-\!3) & : & \mathbf{(15,1) \!
  \oplus \! (1,(n\!-\!3)(2n\!-\!7)) \! \oplus \! (6,2n\!-\!6)}  
\ea
\ed
The table gives the correct results for $4 \geq n > 1$, provided one
restricts to the lines where the number $p$ in the always present
$SO(p,p)$ factor is bigger or equal than $0$ ($SO(0,0)$ is the
trivial group), and omit representations that would be $0$-dimensional
according to the above formula's. 

Notice that $SO(2,2) \cong SL(2,\R)
\times SL(2,\R)$ is not simple. This theory allows an outer
automorphism (T-duality) in 3 dimensions, interchanging the $SL(2,\R)$
of gravity with the $SL(2,\R)$ of S-duality in 4 dimensions. Finally,
$SO(1,1) \cong \R$; this theory cannot be oxidized, because we require a
semi-simple factor.   

Note that again for $n=3$ the ``extra'' line is possible, while the
one in the table for $SL(4,\R)$ is not. This is due to the isomorphism
$SO(3,3) \cong SL(4,\R)$.  Notice that this theory gives pure
gravity, as it should.

\subsection{$E_6$}

The split form of $E_6$ is $E_{6(6)}$. This theory oxidizes to 8
dimensions. Compactifying it on a 2-torus it allows a version of
T-duality acting on two dimensions simultaneously (``M-theory
T-duality'') 

The relevant decompositions are:
\bd
\ba{l@{}cl}
SL(6,\R) \! \times \! SL(2,\R) & : & \mathbf{(35,1) \! \oplus \! (1,3)
  \! \oplus \! (20,2)} \\ 
SL(5,\R) \! \times \! SL(2,\R) \! \times \! \R & : & \mathbf{(24,1)^0 \!
  \oplus \! (1,3)^0  \! \oplus \! (1,1)^0 \oplus }\\  
& & \mathbf{(5,1)^6 \! \oplus \! (\ov{5},1)^{-6} \! \oplus \! (10,2)^3
  \! \oplus \! (\ov{10},2)^{-3}} \\ 
SL(4,\R) \! \times \! SL(2,\R) \! \times \! SL(2,\R) \! \times \! \R &
: & \mathbf{(15,1,1)^0 \! \oplus \! (1,3,1)^0 \! \oplus \! (1,1,3)^0 \!
  \oplus \! (1,1,1)^0   \oplus }\\ 
& & \mathbf{(4,1,2)^3 \! \oplus \! (4,2,1)^{-3} \! \oplus \!
  (\ov{4},1,2)^{-3} \! \oplus \!  (\ov{4},2,1)^3 \oplus }\\ 
& & \mathbf{(6,2,2)^0} \\
SL(3,\R) \! \times \! SL(3,\R)\! \times \! SL(3,\R) & : &
\mathbf{(8,1,1) \! \oplus \! (1,8,1) \! \oplus \! (1,1,8) \oplus }\\ 
& & \mathbf{(3,3,3) \! \oplus \! (\ov{3}, \ov{3}, \ov{3})}\\
SL(2,\R) \! \times \! SL(6,\R) &: & \mathbf{(3,1) \! \oplus \! (1,35)
  \! \oplus \!  (2,20)}\\ 
SL(1,\R) \! \times \! E_{6(6)} & : & \mathbf{78} 
\ea
\ed

\subsection{$E_7$}

The split form of $E_7$ is $E_{7(7)}$. The theory oxidizes to 10
dimensions, where it represents a truncated version of IIB gravity,
but also has a separate branch in 8 dimensions.

The relevant decompositions are:
\bd
\ba{l@{}cl}
SL(8,\R) \! \times \! \{e\} & : & \mathbf{63 \! \oplus \! 70}\\
SL(7,\R) \! \times \! \R  & : & \mathbf{48^0 \! \oplus \! 1^0 \!
  \oplus \! 7^4 \! \oplus \! \ov{7}^{-4} \! \oplus \! 35^2 \! \oplus
  \! \ov{35}^{-2}} \\  
SL(6,\R) \! \times \! SL(2,\R) \! \times \! \R & : & \mathbf{(35,1)^0
  \! \oplus \! (1,3)^0  \! \oplus \! (1,1)^0 \! \oplus \! (6,2)^2 \!
  \oplus \! (\ov{6},2)^{-2} \oplus }\\  
& & \mathbf{(15,1)^{-2} \! \oplus \! (\ov{15},1)^2 \! \oplus \!
  (20,2)^0} \\ 
SL(5,\R) \! \times \! SL(3,\R) \! \times \! \R & : & \mathbf{(24,1)^0
  \! \oplus \! (1,8)^0 \! \oplus \! (1,1)^0 \! \oplus \! (5,1)^{-6} \!
  \oplus  \!(\ov{5},1)^6 \oplus }\\  
& & \mathbf{(5,3)^4 \! \oplus \! (\ov{5}, \ov{3})^{-4} \! \oplus \!
  (10,\ov{3})^{-2} \! \oplus \! (\ov{10},3)^2} \\ 
SL(4,\R) \! \times \! SL(4,\R) \! \times \! SL(2,\R) & : &
\mathbf{(15,1,1) \! \oplus \! (1,15,1) \! \oplus \! (1,1,3) \oplus} \\ 
 & & \mathbf{(4,4,2) \! \oplus \! (\ov{4}, \ov{4},2) \! \oplus \!
  (6,6,1)}\\ 
SL(3,\R) \! \times \! SL(6,\R) & : & \mathbf{(8,1) \! \oplus \! (1,35)
  \! \oplus \! (3,15) \! \oplus \! (\ov{3}, \ov{15})} \\ 
SL(2,\R) \! \times \! SO(6,6) & : & \mathbf{(3,1) \! \oplus \!
  (1,66) \! \oplus \! (2,32)} \\ 
SL(1,\R) \! \times \! E_{7(7)} & : & \mathbf{133}
\ea
\ed

There exists an alternative decomposition with a maximal $SL(6,\R)$
factor  
\bd
\ba{l@{}cl}
SL(6,\R) \! \times \! SL(3,\R) & : & \mathbf{(35,1) \! \oplus \! (1,8)
  \! \oplus \! (15,3) \! \oplus \! (\ov{15}, \ov{3})}  
\ea
\ed

\subsection{$E_8$}

The split form of $E_8$ is $E_{8(8)}$. The 3 dimensional theory was
constructed in \cite{Marcus:1983hb}. This theory oxidizes to the
bosonic sector of 11 dimensional supergravity \cite{Cremmer:1978km}. A
separate branch in 10 dimensions oxidizes to the bosonic sector of IIB
gravity \cite{Schwarz:qr}. This chain includes all maximal
supergravity theories in dimensions $\geq 3$ \cite{Cremmer:1979up}(see
e.g. \cite{Obers:1998fb}). 

The relevant decompositions are:
\bd
\ba{l@{}cl}
SL(9,\R) \! \times \! \{e \} & : & \mathbf{80 \! \oplus \! 84 \!
  \oplus \! \ov{84}} \\ 
SL(8,\R) \! \times \! \R     & : & \mathbf{63^0 \! \oplus \! 1^0 \!
  \oplus \! 8^{-1} 
  \! \oplus \! \ov{8}^1 \! \oplus \! 28^2 \! \oplus \! \ov{28}^{-2} \!
  \oplus \! 56^{-3} \! \oplus \! \ov{56}^3}\\ 
SL(7,\R) \! \times \! SL(2,\R) \! \times \! \R & : & \mathbf{(48,1)^0
  \! \oplus \! (1,3)^0 \! \oplus \! (1,1)^0 \! \oplus \! (7,1)^4 \!
  \oplus \! (\ov{7},1)^{-4} \oplus }\\ 
& & \mathbf{(7,2)^{-3} \! \oplus \! (\ov{7},2)^3 \! \oplus \! (21,2)^1
  \! \oplus \! (\ov{21},2)^{-1} \oplus}\\ 
& & \mathbf{(35,1)^{-2} \! \oplus \! (\ov{35},1)^2}\\
SL(6,\R) \! \times \! SL(3,\R) \! \times \! SL(2,\R) & : &
\mathbf{(35,1,1) \! \oplus \! (1,8,1) \! \oplus \! (1,1,3) \! \oplus
  \! (6,3,2) \oplus }\\   
& & \mathbf{(\ov{6}, \ov{3}, 2)\! \oplus \! (15,\ov{3},1) \! \oplus \!
  (\ov{15},3,1) \! \oplus \! (20,1,2)} \\  
SL(5,\R) \! \times \! SL(5,\R) & : & \mathbf{(24,1) \! \oplus \!
  (1,24) \oplus }\\ 
& & \mathbf{(5,10) \! \oplus \! (10,\ov{5}) \! \oplus \! (\ov{10},5)
  \! \oplus \! (\ov{5},\ov{10})}\\ 
SL(4,\R) \! \times \! SO(5,5) & : & \mathbf{(15,1) \! \oplus \!
  (1,45) \! \oplus \! (4,16) \! \oplus \! (\ov{4}, \ov{16}) \! \oplus
  \! (6,10)}\\ 
SL(3,\R) \! \times \! E_{6(6)} & : & \mathbf{(8,1) \! \oplus \! (1,78)
  \! \oplus \! (3,27) \! \oplus \! (\ov{3}, \ov{27})}\\ 
SL(2,\R) \! \times \! E_{7(7)} & : & \mathbf{(3,1) \! \oplus \!
  (1,133) \! \oplus \! (2,56)}\\ 
SL(1,\R) \! \times \! E_{8(8)} & : & \mathbf{248}\\ 
\ea
\ed

There exist an alternative decomposition with a maximal $SL(8,\R)$
factor. 
\bd
\ba{l@{}cl}
SL(8,\R) \! \times \! SL(2,\R) & : & \mathbf{(63,1) \! \oplus \! (1,3)
  \! \oplus \! (28,2) \! \oplus \! (\ov{28}, 2) \! \oplus \! (70,1)}  
\ea
\ed

\subsection{$F_4$}

The split form of $F_4$ is $F_{4(4)}$. This gives an intriguing theory
in 6 dimensions, discussed in detail in an appendix of
\cite{Cremmer:1999du}. The lower dimensional theories can
be found in \cite{deWit:1992wf}. All allow supersymmetric extensions.

The relevant decompositions are:
\bd
\ba{l@{}cl}
SL(4,\R) \! \times \! SL(2,\R) & : & \mathbf{(15,1) \! \oplus \! (1,3)
  \! \oplus \! (4,2) \! \oplus \! (\ov{4},2) \! \oplus \! (6,3)} \\ 
SL(3,\R) \! \times \! SL(3,\R) & : & \mathbf{(8,1) \! \oplus \! (1,8)
  \! \oplus \! (3, \ov{6}) \! \oplus \! (\ov{3},6)}\\ 
SL(2,\R) \! \times \! Sp(3,\R) & : & \mathbf{(3,1) \! \oplus \! (1,21)
  \! \oplus \! (2,14)} \\ 
SL(1,\R) \! \times \! F_{4(4)} & : & \mathbf{52}
\ea
\ed

\subsection{$G_2$}

The split form of $G_2$ is $G_{2(2)}$. This theory oxidizes to the
bosonic sector of simple
supergravity in 5 dimensions. Details can be found in
\cite{Mizoguchi:1998wv} (see also \cite{deWit:1991nm}).

The relevant decompositions are:
\bd
\ba{l@{}cl}
SL(3,\R) \! \times \! \{e \}  & : & \mathbf{8 \! \oplus \! 3 \! \oplus
  \! \ov{3}} \\ 
SL(2,\R) \! \times \! SL(2,\R)& : & \mathbf{(3,1) \! \oplus \! (2,4)
  \! \oplus \! (1,3)} 
\\ 
SL(1,\R) \! \times \! G_{2(2)}& : & \mathbf{14}
\ea
\ed

\section{Other relevant group theoretic information}\label{details}

\subsection{Maximal compact subgroups for split forms}

The maximal compact subgroups for the split forms are compiled in the
following table:

\bd
\ba{|ccc|}
\hline
 A_n & SL(n,\R) & SO(n) \\ \hline
 B_n & SO(n+1,n) & SO(n+1) \times SO(n) \\ \hline
 C_n & Sp(n,\R) & U(n) \\ \hline
 D_n & SO(n,n) & SO(n) \times SO(n) \\ \hline
 E_6 & E_{6(6)} & Sp(4) \\ \hline
 E_7 & E_{7(7)} & SU(8) \\ \hline
 E_8 & E_{8(8)} & SO(16)\\ \hline
 F_4 & F_{4(4)} & Sp(3) \times SU(2) \\ \hline
 G_2 & G_{2(2)} & SO(4) \\
\hline
\ea
\ed

\subsection{Isomorphism of groups}

In this section we have collected several groups with isomorphic
algebra's. Although their global topology may be different, we have
adapted to the common practice in the physics literature, and
disregarded global differences.

Equivalences for compact groups:
\bd
\ba{c}
 U(1) \cong SO(2) \\
 SU(2) \cong SO(3) \cong Sp(1) \\
 SU(4) \cong SO(6) \\
 Sp(2) \cong SO(5) \\
 SO(4) \cong SU(2) \times SU(2) \\
\ea
\ed

Equivalences for non-compact groups:
\bd
\ba{c}
 \R \cong SO(1,1) \\
 SU(1,1)  \cong SL(2,\R) \cong SO(1,2) \cong Sp(1,\R) \\
 SL(4,\R) \cong SO(3,3) \\
 SU(2,2)  \cong SO(4,2) \\
 SU^*(4)  \cong SO(5,1) \\
 SU(1,3)  \cong SO^*(6) \\
 Sp(1,1)  \cong SO(4,1) \\
 Sp(2,\R) \cong SO(3,2) \\
 SO(2,2)  \cong SL(2,\R) \times SL(2,\R)\\
 SO(1,3)  \cong SL(2,\C) \\
 SO(6,2)  \cong SO^*(8)  \\
\ea 
\ed

\section{The sigma model equations of motion} \label{dereq}

The derivation of the equations and Bianchi identities (\ref{scbi}),
(\ref{bi2}) and (\ref{eqmo2}) is straightforward. Proving that the
sigma model equations of motion are given by (\ref{sceqmo}) however
requires a few subtleties.

Taking the derivative of (\ref{defmin}), and using the Bianchi
identity (\ref{scbi}) and again (\ref{defmin}), one arrives at (all
Greek symbols stand for positive roots, wedges are omitted to
save space)
\bea
\dif F_{(D-1)-\gamma} = \sum_{\alpha-\beta=-\gamma}
N_{\alpha,-\beta} F_{(1)\alpha} F_{(D-2)-\beta} & + & \non \\
\sum_{{\alpha-\beta=-\gamma} \atop {\delta-\epsilon =-\beta }}
  N_{\alpha,-\beta} N_{\delta, -\epsilon} F_{(1)\alpha} 
  F_{(1)\delta}  A_{(D-2)-\epsilon} & - & \hlf
\sum_{{\alpha-\beta=-\gamma} \atop {\delta+\epsilon = \alpha}}
N_{\alpha,-\beta} N_{\delta, \epsilon} F_{(1)\delta}  F_{(1)\epsilon}
A_{(D-2)-\beta} \label{step1}
\eea
The first line of eq. (\ref{step1}) is eq. (\ref{sceqmo}), so we have to
demonstrate that the second line vanishes. Note that it must vanish,
since terms with bare $A_{(D-2)-\beta}$ cannot be gauge invariant.

We proceed with the second line of eq. (\ref{step1}). We eliminate
$\beta$ from the first sum, and rename dummy indices in the second sum:
\be
\sum_{\alpha + \delta-\epsilon=-\gamma}
  N_{\alpha,\delta-\epsilon} N_{\delta, -\epsilon} F_{(1)\alpha} 
  F_{(1)\delta}  A_{(D-2)-\epsilon} - \hlf
\sum_{{\beta-\epsilon=-\gamma} \atop {\alpha+\delta = \beta}}
N_{\beta,-\epsilon} N_{\alpha, \delta} F_{(1)\alpha}  F_{(1)\delta}
A_{(D-2)-\epsilon} 
\ee
To eliminate $\beta$ from the second term, we use (\ref{strcycl}). In
the first term, we antisymmetrize the combination of the two structure
constants in $\alpha$ and $\delta$, and obtain
\be
 \hlf \sum_{\alpha + \delta-\epsilon=-\gamma}
  \left(N_{\alpha,\delta-\epsilon} N_{\delta, -\epsilon} -
  N_{\delta,\alpha-\epsilon} N_{\alpha, -\epsilon} -
N_{-\epsilon,\gamma} N_{\alpha, \delta} \right) F_{(1)\alpha}  F_{(1)\delta}
A_{(D-2)-\epsilon} \label{step3}
\ee
We now study the terms of (\ref{step3}), for fixed $\alpha,\delta,\epsilon$. 
Of course one has $\alpha \neq \delta$, and also $\alpha \neq
-\epsilon$, $\delta \neq -\epsilon$. A possibility is that $\alpha
=\gamma$. In that case one has $\delta-\epsilon = -2 \alpha$ and
$N_{\alpha,\delta-\epsilon}$ vanishes. One can use the antisymmetry of
the structure constants to show that the remaining two terms (with the
same $\alpha,\delta, \epsilon$) cancel against each other. A
similar argument applies if $\delta =\gamma$. The last possibility is
that $\alpha$, $\gamma$, $\delta$ and $-\epsilon$ are all
distinct. Then we use antisymmetry of the structure constants, and
(\ref{strcycl}) to rewrite the relevant terms to
\be
\left(N_{-\epsilon,\delta} N_{\alpha,\gamma} +
  N_{\alpha,-\epsilon} N_{\delta,\gamma} +
 N_{\delta,\alpha} N_{-\epsilon,\gamma} \right) F_{(1)\alpha}  F_{(1)\delta}
A_{(D-2)-\epsilon} \qquad(\alpha + \delta-\epsilon+\gamma=0),
\ee
which vanishes by virtue of eq. (\ref{strjacobi}). This completes the
derivation of eq. (\ref{sceqmo}).

\end{document}